\documentclass[aip,jcp,graphicx,preprint]{revtex4-1}

\usepackage{xcolor}
\usepackage{physics}
\usepackage{multirow}
\usepackage{subcaption}
\usepackage{siunitx}
\usepackage{hyperref}
\usepackage{modiagram}

\usepackage{amsmath}
\usepackage{expl3}
\usepackage{calc}
\usepackage[version=4]{mhchem}

\usepackage{bm}
\usepackage{dirtytalk}
\usepackage{amssymb}
\usepackage{mathrsfs}

\normalsize
\usepackage{booktabs}

\DeclareSIUnit{\atomicunit}{a.u.}
\DeclareSIUnit{\bohr}{\ensuremath{\mathit{a}_0}}
\DeclareSIUnit\angstrom{\text{Å}}
\DeclareSIUnit\barn{b}
\AtBeginDocument{\RenewCommandCopy\qty\SI}

\begin{document}

\title{Influence of vibrational motion and temperature on interatomic Coulombic electron capture} 

\author{Elena M. Jahr}
    \affiliation{Institute of Physical and Theoretical Chemistry, University of Tübingen, Auf der Morgenstelle 18, 72076 Tübingen, Germany}
    \affiliation{Center for Light-Matter Interaction, Sensors and Analytics (LISA+), University of Tübingen, Auf der Morgenstelle 15, 72076 Tübingen, Germany}
\author{Jan \v{S}enk}
    \affiliation{Laboratoire de Chimie Physique–Matière et Rayonnement, UMR 7614, Sorbonne Université, CNRS, F-75231 Paris Cedex 05, France}
    \affiliation{Institute of Theoretical Physics, Faculty of Mathematics and Physics, Charles University, V Holešovičkách 2, 180 00 Prague, Czech Republic}
\author{Jan P. Drennhaus}
    \affiliation{Institute for Theoretical Physics, Department of Physics and Astronomy, KU Leuven, Celestijnenlaan 200d, box 2415, 3001 Leuven, Belgium}
\author{P\v{r}emysl Koloren\v{c}}
    \affiliation{Institute of Theoretical Physics, Faculty of Mathematics and Physics, Charles University, V Holešovičkách 2, 180 00 Prague, Czech Republic}
\author{Nicolas Sisourat}
    \affiliation{Laboratoire de Chimie Physique–Matière et Rayonnement, UMR 7614, Sorbonne Université, CNRS, F-75231 Paris Cedex 05, France}
\author{Elke Fasshauer}
    \email{elke.fasshauer@gmail.com}
    \affiliation{Institute of Physical and Theoretical Chemistry, University of Tübingen, Auf der Morgenstelle 18, 72076 Tübingen, Germany}
    \affiliation{Center for Light-Matter Interaction, Sensors and Analytics (LISA+), University of Tübingen, Auf der Morgenstelle 15, 72076 Tübingen, Germany}

\date{\today}

\begin{abstract}
Interatomic Coulombic Electron Capture (ICEC) is an environment-mediated process in which a free electron attaches to a species by transferring excess energy to a neighbor.
While previous theoretical investigations assumed fixed nuclei, recent studies indicate that nuclear dynamics significantly influences the ICEC process.
In this work, we incorporate the vibrational motion into an analytical model of the ICEC cross section including both energy and electron transfer.
To validate this approach, we compare the results to the adiabatic-nuclei approximation based on fixed-nuclei \textit{ab initio} R-matrix calculations.
We apply our theory to the helium-neon dimer, which is ideal for studying diverse dynamical effects.
We show that while vibrational dynamics can slightly reduce ICEC efficiency, ICEC remains dominant over photorecombination and can trigger dimer dissociation.
Accounting for the nuclear motion also enables to describe the broadening of the electron spectrum and enables evaluation of temperature-dependent cross sections -- capabilities beyond the reach of fixed-nuclei approaches.
\end{abstract}

\pacs{}% insert suggested PACS numbers in braces on next line

\maketitle

\section{Introduction}

The capture of free electrons by atoms, ions, molecules, and quantum dots is a fundamental process with wide-ranging implications in both basic and applied sciences. 
It plays a key role in astrophysics and plasma physics, where it influences stellar processes and ionized gases~\cite{tucker, astro}. 
Furthermore, electron capture is critical for understanding electrochemical and photochemical reactions, impacting biological systems~\cite{dna} and solid-state technologies~\cite{semiconductors}.

One such electron capture mechanism is the Interparticle Coulombic Electron Capture (ICEC) that occurs exclusively in systems embedded within an environment~\cite{gokhberg2010}.
In ICEC, a free electron is captured by an electron acceptor (\ce{A}), which in this work will be a cation, while the excess energy is transferred via Coulomb interaction to a neighboring electron donor (D), leading to the ionization of D:
\begin{equation}
e^-_{k} + \ce{A} + \ce{D} \to \ce{A-} + \ce{D+} + e^-_{k'}.
\label{eq:reaction}
\end{equation}
This environment-mediated coupling significantly enhances the probability of electron capture across various systems, as shown in theoretical studies~\cite{gokhberg2009, molle_water-assisted_2023}. 
A detailed overview of the mechanism and its theoretical basis can be found in Ref.~\onlinecite{bande_2023}.

In most earlier studies, theoretical approaches to investige ICEC have assumed static nuclei, neglecting the influence of nuclear dynamics~\cite{bande_2023}. 
However, recent results~\cite{fede_2024} reveal that nuclear motion, particularly vibrational dynamics, can substantially affect the efficiency and outcome of the process. 
Findings for related processes, such as resonant two-center photoionization~\cite{gruell_2020} and two-center dielectronic recombination~\cite{jan_2025}, further support the importance of nuclear motion in interparticle electron capture. 
These findings highlight the necessity of incorporating nuclear dynamics into ICEC models to achieve a more complete theoretical understanding.

In this work, we extend the previous analytical energy and electron transfer model~\cite{gokhberg2009, senk2024} of ICEC by explicitly incorporating the relative nuclear motion between the interacting partners, 
%In this work, we extend an analytical model of ICEC by explicitly incorporating the relative nuclear motion between the interacting partners. 
%We include both the energy transfer~\cite{gokhberg2009} and the electron transfer~\cite{senk2024} contributions, 
providing intuitive insights grounded in the physical characteristics of the individual systems.
%Throughout this paper, we will reference this model as analytical or energy and electron transfer model of ICEC.
We apply this approach to the positively charged helium–neon dimer and benchmark our results against the adiabatic-nuclei approximation based on fixed-nuclei \textit{ab initio} R-matrix calculations, which offer a rigorous quantum mechanical treatment of ICEC as a scattering process~\cite{sisourat_2018, molle_2021}. 
Using both approaches, we assess the impact of nuclear dynamics on ICEC and aim to provide theoretical guidance for future experimental efforts.

The paper is organized as follows: 
We first establish the theoretical framework in Sec.~\ref{sec:theory}, incorporating nuclear degrees of freedom into the analytical model of ICEC.
In Sec.~\ref{sec:methods_R-matrix}, we describe the approximations taken to include nuclear dynamics into the R-matrix approach.
% we describe the fixed-nuclei R-matrix method and the approximations taken to include nuclear dynamics into this framework.
Section~\ref{sec:system} describes the physical parameters of our example of a weakly bound system, the positively charged helium-neon dimer. 
Section~\ref{sec:computational_details} outlines the computational details. 
Results for the ICEC cross sections of the helium-neon dimer are discussed in Sec.~\ref{sec:results}. 
Finally, we summarize our findings and their implications in Sec.~\ref{sec:conclusion}. 
We use atomic units ($\hbar = m_e = e = 1$) throughout this paper, unless explicitly mentioned otherwise.

%We then apply this approach to the positively charged helium-neon dimer as a model for a weakly bound system to analyze its impact on electron capture efficiency.
%The paper is organized as follows: 
%Section II describes the theoretical approaches employed in this study. 

%First we focus on the asymptotic approximation which describes the process with an analytical formula based on physical properties of the individual species \cite{gokhberg2009, gokhberg2010}.

%In the R-matrix framework, we apply a version of the adiabatic-nuclei approximation (to be cited) to incorporate vibrational dynamics using fixed-nuclei R-matrix results. 

\iffalse
For weak interactions between ICEC partners, an analytical formula for the ICEC cross sections can be derived, referred to as the asymptotic approximation. 
Cross sections are then estimated based on physical properties of the individual species \cite{gokhberg2009, gokhberg2010}.
For the asymptotic approach, we utilize analytical solutions based on Morse potentials to account for the vibrational motion. 

ICEC has also been explored using the \textit{ab initio} fixed-nuclei R-matrix method \cite{sisourat_2018, molle_2021}, 
\fi

%\input{theory}

\section{Nuclear dynamics in the analytical energy and electron transfer model of ICEC}
\label{sec:theory}
From a scattering theory perspective, ICEC is a multichannel inelastic process in which an incoming electron interacts with a target~\cite{gokhberg2009, taylor2006}.
It is characterized by a cross section that quantifies its probability for a given electron energy.
Our goal is to derive an approximation to the ICEC cross section that includes nuclear dynamics.

We consider the process in Eq.~\eqref{eq:reaction}, where an electron is captured by an electron acceptor A, and a neighboring electron donor D emits an electron.
Both species can be atoms, molecules, or ions.
The corresponding differential cross section is
\begin{equation}
    \label{eq:diff_xs}
    \dv{\sigma}{\Omega_{\mathbf{k}'}} 
    = \frac{1}{(2\pi)^2}\frac{k'}{k} |M|^2 ,
\end{equation}
where $\Omega_{\mathbf{k}'}$ is the solid angle of the outgoing electron and $M$ is the transition matrix element~\cite{gokhberg2009, taylor2006}, in our case between two vibronic states $\Psi_\alpha$ accounting for the vibrational
dynamics between A and D.
%The electronic wave function of the system is antisymmetrized to satisfy the Pauli exclusion principle. 
We describe the vibronic states as a single term of the Born-Huang expansion \cite{Ballhausen1972, Born1996}: 
\begin{equation}
\label{eq:vibronic_state}
    |\Psi_{\alpha} (\mathbf{r},R) \rangle = |\phi_\alpha(\mathbf{r}\,;R)\rangle \, |\nu_\alpha(R)\rangle,
\end{equation}
where $\phi_\alpha$ is the electronic wave function, $\nu_\alpha$ is the nuclear wave function, $\mathbf{r}$ are the electronic coordinates, and $R$ is the distance between the centers of mass of A and D.

As shown in Ref.~\onlinecite{senk2024}, the antisymmetry of the electronic wave function leads to two distinct contributions to $M$~\cite{taylor2006}: 
\begin{equation}
    M(\mathbf{k}' \leftarrow \mathbf{k}) 
    = M_\text{el}(\mathbf{k}' \leftarrow \mathbf{k}, e^-_{k}=e^-_{k'}) + M_\text{en}(\mathbf{k}' \leftarrow \mathbf{k}, e^-_{k}\neq e^-_{k'}).
\end{equation}
The electron transfer term, $M_\text{el}$, assumes the incoming and outgoing electron are the same and describes direct scattering accompanied by electron relocation from donor to acceptor.
The energy transfer term, $M_\text{en}$, involves capture of the incoming electron by A, with excess energy transferred to D, which emits an electron.

These two mechanisms dominate at different distances between the two particles: electron transfer requires short-range interactions, while energy transfer remains effective even at larger separations~\cite{graves_2024}.
In the following, we focus on incorporating nuclear dynamics into the energy transfer contribution before turning to electron transfer.

% ===============================

\subsection{Energy transfer contribution to ICEC}

For the energy transfer contribution $M_\text{en}$, we extend the derivation from Refs.~\onlinecite{gokhberg2010,gokhberg2009}, incorporating the necessary modifications to account for nuclear motion.
Here, we treat the process as events taking place separately on A and D, linked by the energy transfer via first order of perturbation theory. 
This approach is referred to as "asymptotic" or "virtual photon" approximation in the literature~\cite{bande_2023}.
We will see that this enables us to express the cross section in terms of physical parameters of A and D~\cite{gokhberg2009, gokhberg2010}, which can be found in literature or obtained from experiments.

We begin by approximating the initial (i) and final (f) electronic wave functions as products of the individual states of A and D,
\begin{equation}
\begin{aligned}
\label{eq:el_states}
    \ket{\phi_i} &= \ket{\phi_{A, \mathbf{k}}} \ket{\phi_{D}}, \\
    \ket{\phi_f} &= \ket{\phi_{A^-}} \ket{\phi_{D^+, \mathbf{k'}}}, \\
\end{aligned}
\end{equation}
where the energy-normalized state $\phi_{A, \mathbf{k}}$ ($\phi_{D^+, \mathbf{k'}}$) describes A (\ce{D+}) together with the incoming (outgoing) electron of momentum $\mathbf{k}$ ($\mathbf{k}$')~\cite{gokhberg2010}.
This is justified if A and D are far enough away from each other and the electronic interaction between them can be assumed to be weak.

By using Eqs.~\eqref{eq:el_states} and \eqref{eq:vibronic_state} and employing the Born-Oppenheimer approximation, the transition matrix element within the first order of perturbation theory for a single initial and final state is
\begin{equation}
    \label{eq:Men}
    M_\text{en} 
    = \bra{\nu_f} \bra{\phi_{A^-}} \mel{\phi_{D^+,\mathbf{k}'}}{\hat{V}}{\phi_{A,\mathbf{k}}} \ket{\phi_D} \ket{\nu_i},
\end{equation}
where $\hat{V}$ is the Coulomb interaction between the charges of the two systems. 
%Note that $\nu_i$ ($\nu_f$) is a vibrational eigenstate of a potential energy surface where the particles are separated as $A^+$ and $D$ ($A$ and $D^+$) in the $R\to \infty$ limit.  

For sufficiently large distance $R$ between A and D, we expand $\hat{V}$ in a Taylor series around $R$ and retain the leading term that yields a non-zero contribution to the transition matrix element $M_\text{en}$.
This term corresponds to the dipole-dipole interaction of the two charge densities
\begin{equation}
    \label{eq:Vmumu}
    \hat{V}_{\mu\mu} 
    = \frac{1}{R^3} \, \big(\hat{\boldsymbol{\mu}}_A \cdot  \hat{\boldsymbol{\mu}}_D 
    - 3 \, (\mathbf{u}_R \cdot \hat{\boldsymbol{\mu}}_A) \, (\mathbf{u}_R \cdot \hat{\boldsymbol{\mu}}_D) \big),
\end{equation}
where $\mathbf{u}_R$ is the unit vector pointing from A into the direction of D.
%in the direction of the inter-particle axis.
As the electronic dipole moment operators $\hat{\boldsymbol{\mu}}_A$ and $\hat{\boldsymbol{\mu}}_D$ of A and D are independent of $R$, we can separate the integration over electronic and nuclear coordinates.

Substituting Eqs.~\eqref{eq:Men} and \eqref{eq:Vmumu} into Eq.~\eqref{eq:diff_xs} yields the differential cross section in the lowest-order approximation for the energy transfer contribution to ICEC:
%As a result, the transition matrix element is
%\begin{equation}
    %M_\text{en} = 
    %\big( \boldsymbol{\mu}_A \cdot  \boldsymbol{\mu}_D - 3 \, (\mathbf{u}_R \cdot \boldsymbol{\mu}_A) \, (\mathbf{u}_R \cdot \boldsymbol{\mu}_D) \big)
    %\cdot \langle \nu_f| R^{-3} | \nu_i \rangle, 
%\end{equation}
%The differential cross section then reads
\begin{equation}
    \label{eq:dif_XS_energy_tf}
    \dv{\sigma}{\Omega_{\mathbf{k}'}} 
    = \frac{1}{(2\pi)^2}\frac{k'}{k} 
    \left| \Big[
    \boldsymbol{\mu}_A \cdot  \boldsymbol{\mu}_D - 3 \, (\mathbf{u}_R \cdot \boldsymbol{\mu}_A) \, (\mathbf{u}_R \cdot \boldsymbol{\mu}_D) 
    \Big]
    \left\langle \nu_f \left|  R^{-3} \right| \nu_i \right\rangle \right|^2,
\end{equation}
where $\boldsymbol{\mu}_A = \langle \phi_{A^-} | \, \hat{\boldsymbol{\mu}}_A \, | \phi_{A, \mathbf{k}} \rangle$ and $\boldsymbol{\mu}_D= \langle \phi_{D^+, \mathbf{k}'} | \, \hat{\boldsymbol{\mu}}_D \, | \phi_{D} \rangle$ are the electronic transition dipole moments.
To obtain the total cross section, Eq.~\eqref{eq:dif_XS_energy_tf} is integrated over the solid angle $\Omega_{\mathbf{k}'}$ of the outgoing electron and averaged over the incoming electron angle $\Omega_{\mathbf{k}}$.
To facilitate this integration, one expands the plane waves in the initial and final states in terms of spherical waves~\cite{gokhberg2010, sobelman_1973}.
%We associate the photoionization (PI) and photorecombination (PR) cross section with the corresponding transition dipole moments

The dipole transition moments involved in the process can be related to the corresponding photorecombination (PR) and photoionization (PI) cross sections.
The PR cross section of the electron acceptor A is given by~\cite{sobelman_1973}
\begin{equation}
    \label{eq:photorecombination}
    \sigma^\mathrm{PR}_{A}(\varepsilon)
    =  \frac{4\pi^2 \omega^3}{3 c^3 k^2} \, 
    |\boldsymbol{\mu}_A|^2,
\end{equation}
where $c$ is the speed of light and $\varepsilon$ is the kinetic energy of the incoming electron, corresponding to momentum $k$.
The quantity $\omega$ is the energy of the emitted photon, corresponding here to the excess energy of the electron capture.
Its value is given by the sum of $\varepsilon$ and the electron attachment energy of A, which is equal to the ionization potential of \ce{A-}, $\text{IP}_{\text{A}^-}$.

Similarly, the PI cross section for the electron donor D is~\cite{sobelman_1973}
\begin{equation}
     \sigma^{\text{PI}}_D(\omega') 
    = \frac{4 \pi^2 \omega'}{3 c} \,  
     |\boldsymbol{\mu}_D|^2.
\end{equation}
where $\omega'$ corresponds to the energy available to ionize D.
The difference of $\omega'$ and $\omega$ is $\Delta \omega_{fi} = E_f - E_i$, where $E_i$ ($E_f$) is the energy of the initial (final) vibrational state of the dimer.
%with respect to the corresponding asymptotic energy at $R\to\infty$.
As a result, the kinetic energy of the outgoing electron is
\begin{equation}
    \varepsilon' =  \omega - \Delta\omega_{fi} - \text{IP}_\text{D}.
    \label{eq:energy_conservation_A}
\end{equation}
where $\text{IP}_\text{D}$ is the ionization potential of the electron donor D.

Using these two expressions, we arrive at an approximation to the energy-transfer contribution to the ICEC cross section for the transition from vibrational state $|\nu_i\rangle$ of \ce{AD} to vibrational state $|\nu_f\rangle$ of \ce{A-}\ce{D+}
\begin{equation}
    \label{eq:en_transfer}
    \sigma (k, \nu_i \to \nu_f) 
    = \frac{3 c^4 }{4 \pi} \,
    \frac{\sigma^\text{PR}_{A}(\varepsilon) \, \sigma^\text{PI}_D(\omega - \Delta \omega_{fi})}
        {\omega^3(\omega - \Delta \omega_{fi})} \,
    \left|    
    \bra{\nu_f} R^{-3} \ket{\nu_i}
    \right|^2.
\end{equation}

%The transferred energy $\omega = \varepsilon + \text{IP}_{A}$ is the sum of the kinetic energy $\varepsilon$ of the incoming electron and the ionization potential (IP) of A.

Having established a working expression for the energy transfer contribution to ICEC, we now turn our attention to the electron transfer mechanism.

%We can further substitute the PR cross section by the corresponding PI cross section using the principle of detailed balance~\cite{sobelman_1973}
%\begin{equation}
    %\sigma (k, \nu_i \to \nu_f) 
    %= \frac{3 c^2 }{8 \pi} \,
    %\frac{g_{A^-}}{g_A} \,
    %\frac{\sigma_\text{PI}^A(\omega) \, \sigma_\text{PI}^D(\omega - \Delta \omega_{fi})}
        %{\varepsilon \, \omega \, (\omega - \Delta \omega_{fi})} \,
    %\left|    
    %\bra{\nu_f} R^{-3} \ket{\nu_i} 
    %\right|^2
%\end{equation}

% ===================================

\subsection{Electron transfer contribution to ICEC}
For the electron transfer contribution $M_\text{el}$, we extend the derivation from Ref.~\onlinecite{senk2024}, modifying to account for nuclear motion, analogously to the previous section.
In this approach, the overlap of the electronic wave functions is not neglected, but instead approximated by the overlap of two Gaussians~\cite{mahanty_1973, franz_2921, niggas_peeling_2021, senk2024}.
%This idea stems from the Gaussian approximation to the polarizability in the related interatomic Coulombic decay used in .

In the electron transfer contribution, we assume that we can distinguish the scattering electron from the bound electrons of \ce{(AD)}.
We further assume that we can treat the interaction of the continuum electron and the target as a simple interaction of charge distributions, neglecting electron correlation.
We can then write our initial and final electronic states as
\begin{equation}
\begin{aligned}
\label{eq:el_states_2}
    |\phi_i\rangle &= |\phi_{AD}\rangle |\mathbf{k} +\rangle, \\
    |\phi_f\rangle &= |\phi_{A^-D^+}\rangle |\mathbf{k}' -\rangle,
\end{aligned} 
\end{equation}
where $\phi_{AD}$ ($\phi_{A^-D^+}$) is the electronic wave function of the dimer and $\mathbf{k}$ ($\mathbf{k}'$) is the initial (final) momentum vector of the continuum electron.
We denote with $+$ ($-$) the incoming (outgoing) scattering state~\cite{taylor2006} where the continuum electron scatters off of \ce{AD} (\ce{A-}\ce{D+}).

Although we do not consider the electrons of A and D as distinguishable, the electron correlation between them can still be treated as a perturbation.
Applying Eqs.~\eqref{eq:vibronic_state} and~\eqref{eq:el_states_2} within the Born-Oppenheimer approximation, the transition matrix element at first order of perturbation theory for a single initial and final state is given by
\begin{equation}
    M_\text{el} 
    =  \big\langle  \nu_f \big| \, \langle \mathbf{k}' -  | 
    \langle \phi_{A^-D^+} |\, \hat{V} \, |\phi_{AD} \rangle | \mathbf{k}+ \rangle
     \,
    \big| \nu_i \big\rangle,
\end{equation}
where $\hat{V}$ is the Coulomb interaction between electrons situated on A (including the incoming electron) and on D, and $\phi$ now neglects the correlation between electrons on A and D as the zero-order wave function.
%We will not consider higher order terms in this paper.

%Next, we include the vibrational states in a product ansatz, employing the Born-Oppenheimer approximation as before.
We can again apply the multipole expansion to $\hat{V}$ (cf.~Eq.~\eqref{eq:Vmumu}), but the lowest contributing order is now the $R^{-1}$ term, as we assume the wavefunctions of A and D to have a nonzero overlap.
We obtain the transition matrix element as
\begin{equation}
\label{eq:el_transitionmatrix}
    M_\text{el} 
    = \big\langle  \nu_f \big| \, 
    \frac{\langle \phi_{A^-D^+} | \phi_{AD} \rangle \,
    \langle \mathbf{k}' - | \mathbf{k}+ \rangle}
    {R} \,
    \big| \nu_i \big\rangle.
\end{equation}
%\textcolor{red}{include $\langle \phi_{AD^+} |\phi_{A^+D}\rangle  \, \langle k' -| k + \rangle$ ? Is this product ansatz?} 
%The integral over the electronic coordinates, can be separated into the overlap of the initial and final electronic wave functions of the dimer $S_{AD^+}$ and the overlap the incoming and outgoing electron $\langle k' -| k + \rangle$ 
%\begin{equation}
%     \langle \phi_{AD^+},\mathbf{k}' - | \phi_{A^+D} , \mathbf{k}+ \rangle = S_{AD}(R) \, \langle k' -| k + \rangle.
%\end{equation}
%where $|k +\rangle$ ($|k -\rangle$) is the incoming (outgoing) scattering state of the continuum electron.
%\textcolor{red}{what exactly is k+, k-?}

Assuming a single Slater determinant for the initial and final electronic state of the dimer, the overlap $\langle \phi_{A^-D^+} | \phi_{AD}\rangle$ reduces to the product of the virtual orbital of A, which accepts an electron, $\phi^A_\alpha$, and the orbital of D, which donates said electron, $\phi^D_\delta$.
Approximating these two orbitals as Gaussians of widths $a_A$ and $a_D$ results in the following expression for the overlap~\cite{senk2024}
\begin{equation}
\label{eq:S_AD}
    S_{AD}(R) 
    = \langle \phi^A_\alpha| \phi^D_\delta \rangle \\
    = \left( \frac{a_A a_D}{a_A^2 + a_D^2} \right)^{3/2} \exp \left( - \frac{R^2}{2 (a_A^2 + a_D^2)}\right).
\end{equation}
Note that the true overlap of the two orbitals can vanish due to their symmetries.

Now, let us gather the results for the approximation to the differential cross section of the electron transfer path of ICEC by substituting Eq.~\eqref{eq:el_transitionmatrix} into Eq.~\eqref{eq:diff_xs},
\begin{equation}
\label{eq:el_diffXS}
    \dv{\sigma}{\Omega_{\mathbf{k}'}} 
    = \frac{1}{(2\pi)^2}\frac{k'}{k} \, 
    \Big| \big\langle  \nu_f \big| \, 
    \frac{ S_{AD}  \,  \langle \mathbf{k}' - | \mathbf{k}+ \rangle }{R}
    \big| \nu_i \big\rangle \Big|^2.
\end{equation}

To obtain the total cross section $\sigma$, we integrate Eq.~\eqref{eq:el_diffXS} over the solid angle $\Omega_{\mathbf{k}'}$ of the outgoing electron. 
The angular dependence enters only through the overlap of the incoming and outgoing scattering states, $\langle \mathbf{k}' - | \mathbf{k}+ \rangle$. 
Assuming the potential of the dimer can be approximated by its spherical component, we employ a partial wave expansion of the scattering states~\cite{taylor2006}, leading to
\begin{equation}
    \label{eq:partial_wave_expansion}
    \langle \mathbf{k}' - | \mathbf{k} + \rangle = \frac{1}{2 \pi^2 k k'} 
    \int_0^\infty \dd{r}
    \sum_{\ell = 0}^\infty (2\ell+1) \,
    \psi_{\ell \, \mathbf{k}'}^{- \, *}(r,R) \, \psi_{\ell \, \mathbf{k}}^+(r,R) \,
    P_\ell(\hat{\mathbf{k}}'\cdot \hat{\mathbf{k}}),
\end{equation}
where the hat denotes unit vectors, $\psi^\pm$ are the radial incoming and outgoing scattering wave functions, and $P_\ell$ is the Legendre polynomial.
The angular parts of this expansion -- the Legendre polynomials -- do not depend on $R$, hence we can carry out the integration over $\Omega_{\mathbf{k}'}$.
%We can interchange scalar terms and then integrate the angular part of $\langle \mathbf{k}' - | \mathbf{k}+ \rangle$, i.e. the Legendre Polynomials, first over $\Omega_{\mathbf{k}'}$.
If we then define
\begin{equation}
    J_\ell := \int_0^\infty \dd{r} \psi_{\ell \, \mathbf{k}'}^{- \, *}(r,R) \, \psi_{\ell \, \mathbf{k}}^+(r,R),
\end{equation}
%\begin{equation}
%    \int \dd{\Omega_{\mathbf{k}'}} \, |  \langle \mathbf{k}' - | \mathbf{k}+ \rangle |^2
%    = \frac{1}{\pi^2 k^2 k'^2} \sum_{\ell=0}^\infty (2\ell + 1) J^2_\ell
%\end{equation}
%where $J_\ell$ is the overlap of the radial incoming and outgoing scattering $\ell$-type wave functions of the continuum electron.
% $J_\ell$ is defined as the square of the definition in the present work and that
the total cross section reads
\begin{equation}
    \label{eq:el_transfer}
    \sigma_\text{el} = 
    32 \pi \, \frac{1}{\sqrt{\varepsilon^3 \varepsilon'}} \,
    \sum_{\ell=0}^\infty (2\ell + 1) \,
    \Big| \big\langle \nu_f \big| \, S_{AD} \, J_\ell \, R^{-1} \, \big| \nu_i\big\rangle \Big|^2,
\end{equation}
where $\varepsilon$ ($\varepsilon'$) is the kinetic energy of the incoming (outgoing) electron.
Note that compared to Ref.~\onlinecite{senk2024}, the integral defining $J_\ell$ is not squared and that we cannot simplify the absolute square further due to the integration over $R$.

We further assume an exponential decrease of $J_\ell$ squared, as in Ref.~\onlinecite{senk2024},
\begin{equation}
     J_\ell^2 \approx C \exp\left( - \frac{\ell(\ell + 1)}{ K^2_{AV}}\right),  
\end{equation}
where we defined $K^2_{AV} = (a_A + R)^2 \varepsilon + (a_D + R)^2 \varepsilon'$, and an ad hoc dependence of $C$ on $\varepsilon'-\varepsilon$
\begin{equation}
    C \approx \Bar{C} \exp\left( - \frac{|\varepsilon'-\varepsilon|}{d}\right),
\end{equation}
where $\Bar{C}$ and $d$ are fitting parameters taken from Ref.~\onlinecite{senk2024}.

With both the energy and electron transfer contributions established for a single vibrational transition, let us summarize the results. 

\subsection{Final considerations of the analytical energy and electron transfer model}
\label{sec:theory_considerations}
In our model, the total ICEC cross section is given by the sum of energy and electron transfer contributions,
\begin{equation}
    \sigma = \sigma_\text{en} + \sigma_\text{el},
\end{equation}
with expressions provided in Eqs.~\eqref{eq:en_transfer} and \eqref{eq:el_transfer}, respectively.
Although these mechanisms could, in principle, interfere, their transition amplitudes are only comparable over a narrow range of $R$~\cite{senk2024}. 
Since the integration spans all internuclear distances, their direct interference contributes only marginally, and an incoherent sum of cross sections is appropriate.
%Here, we additionally integrate over $R$, so in principle, only a small part of the integrand is comparable in value.

So far, we have considered only a single vibrational transition.
In practice, ICEC involves transitions to multiple bound and dissociative final states. For the former, we sum over all final vibrational states $\nu_f$, and for the latter, we integrate over the dissociation energy $E$:
\begin{equation}
    \label{eq:sum_over_v}
    \sigma (k, \nu_i) 
    = \sum_f \sigma (k, \nu_i \to \nu_f) 
    + \int_{E=0}^{E_\text{max}} \dv{\sigma}{E} {(k, \nu_i \rightarrow E)} \, \dd{E},
\end{equation}
where $E_\text{max} = \omega - \text{IP}_D + E_i$ is the maximum dissociation energy allowed by energy conservation, corresponding to the limit of $\varepsilon'=0$. 

To account for a thermal population in the initial state, we average over initial vibrational states using a Boltzmann distribution, which introduces temperature dependence:
\begin{equation}
    \label{eq:boltzmann}
    \sigma (k, T) 
    = \frac{1}{\sum_i e^{-E_i/(k_B T)}} \sum_i  e^{-E_i/(k_B T)} \sigma (k, \nu_i),
\end{equation}
where $E_i$ is the energy of vibrational state $\nu_i$, $T$ is the thermodynamic temperature, and $k_B$ is the Boltzmann constant.
This temperature dependence is a key feature of our model, made possible by explicitly including nuclear motion.
%This temperature dependence was absent in previous treatments that considered only static nuclei, highlighting the crucial role of nuclear dynamics in a more comprehensive description of the process.

%Note that, as $v=0$ is a bound state, its energy is negative (w.r.t. to the dissociation limit).

%If we further consider multiple initial (final) electronic states of the dimer, we average (sum) over them, taking into account their degeneracy
%\begin{equation}
%    \sigma (k) = \frac{1}{g_\alpha} \sum_\alpha \sum_\beta \sigma (k, \alpha\to \beta),
%\end{equation}
%where $\alpha$ and $\beta$ represent the different electronic states.

With this theoretical foundation, we now turn to the R-matrix approach.
%Having established the lowest-order model, we now turn to the R-matrix method. 

%\input{methods}
\section{Nuclear dynamics in the R-matrix approach}
\label{sec:methods_R-matrix}

We aim to compare the results of the semi-empirical model introduced in Sec.~\ref{sec:theory} with a more rigorous approach. We calculate the cross section for the vibronic transitions using an approximation building upon the fixed-nuclei scattering T-matrix. We evaluate this T-matrix using the \textit{ab initio} R-matrix method~\cite{rmatrix,tennyson_2010}, which is well suited for low-energy electron-molecule scattering calculations. 
The fixed-nuclei cross section is given in terms of the T-matrix as~\cite{tennyson_2010}
\begin{align}
    \sigma_{i \rightarrow f} (\varepsilon;R) = \frac{\pi}{k^2} \sum_{l m l^\prime m^\prime} |T_{i l m \rightarrow f l^\prime m^\prime}(\varepsilon,\varepsilon^\prime;R)|^2,
    \label{eq:XS_in_terms_of_T}
\end{align}
where $l,m$ ($l^\prime,m^\prime$) denote all the contributing partial waves of the incoming (outgoing) electron.

We employ the adiabatic-nuclei (AN) approximation~\cite{chase_adiabatic_1956} to compute the vibrationally resolved scattering cross section. 
In this approach, the vibrational scattering matrix is constructed as follows
\begin{align}
    T_{i \nu_i \rightarrow f \nu_f} (\varepsilon,\varepsilon^\prime) = \mel{\nu_f}{T_{i \rightarrow f} (\varepsilon,\tilde{\varepsilon}^\prime;R)}{\nu_i}_R.
    \label{eq:AN}
\end{align}
The incoming $\varepsilon$ and the outgoing $\varepsilon^\prime$ electron energies in the vibrationally resolved picture are related by the conservation of energy (an equivalent of Eq.~\eqref{eq:energy_conservation_A})
\begin{align}
    \varepsilon + \mathscr{E}_i^{\nu_i} = \varepsilon^\prime + \mathscr{E}_f^{\nu_f},
\end{align}
where $\mathscr{E}_i^{\nu_i}$ ($\mathscr{E}_f^{\nu_f}$) is the energy of the initial (final) vibronic state. If we define the energy reference as the energy of the final electronic state in the limit of large $R$, then $\mathscr{E}_f^{\nu_f} = E_f$, which equals the kinetic energy release (KER). 
The approximation utilizes the fixed-nuclei on-shell T-matrix, $T_{i \rightarrow j} (\varepsilon,\tilde{\varepsilon}^\prime;R)$, where the incoming $\varepsilon$ and the outgoing $\tilde{\varepsilon}^\prime$ electron energies are related by the $R$-dependent conservation of energy 
\begin{align}
    \varepsilon + E_i (R) = \tilde{\varepsilon}^\prime + E_f (R),
    \label{eq:AN_fixed_nuclei_energy_conservation}
\end{align}
where $E_i (R)$ ($E_f (R)$) is the potential energy curve of the initial (final) electronic state. This means that $\tilde{\varepsilon}^\prime \neq \varepsilon^\prime$, i.e., the cross section for a vibronic transition at given outgoing electron energy $\varepsilon^\prime$ is calculated from fixed-nuclei data corresponding to a different, in fact $R$-dependent, outgoing energy $\tilde{\varepsilon}^\prime$. 
This is because the AN approximation does not take the energy of the vibrational states into account. 
Nevertheless, it has been used successfully for vibrational~\cite{faisal_application_1972}, rotational~\cite{temkin_rotational-vibrational_1974}, and electronic~\cite{shugard_theory_1975} excitations and it is valid for electron energies larger than the corresponding excitation thresholds~\cite{chase_adiabatic_1956}. 

In ICEC, dissociation in the final state can be prominent, making the adiabatic-nuclei approximation invalid in a larger range of electron energies above the ICEC threshold. To remedy this, we use the energy-balanced adiabatic-nuclei (EBAN) approximation introduced for calculating electron impact dissociation of $\text{H}_2$ in Ref.~\onlinecite{stibbe_near-threshold_1998}. In this approximation, we use 
\begin{align}
    T_{i \nu_i \rightarrow f \nu_f} (\varepsilon,\varepsilon^\prime) = \mel{\nu_f}{T_{i \rightarrow f} (\tilde{\varepsilon},\varepsilon^\prime;R)}{\nu_i}_R,
    %\tag{\ref{eq:AN}$^\prime$}
    \label{eq:EB_AN}
\end{align}
instead of Eq.~\eqref{eq:AN}, and
\begin{align}
    \tilde{\varepsilon} + E_i (R) = \varepsilon^\prime + E_f (R),
    %\tag{\ref{eq:AN_fixed_nuclei_energy_conservation}$^\prime$}
    \label{eq:EBAN_fixed_nuclei_energy_conservation}
\end{align}
instead of Eq.~\eqref{eq:AN_fixed_nuclei_energy_conservation}. In this variant, the vibronic outgoing electron energy coincides with the fixed-nuclei outgoing electron energy, and the non-equivalence is transferred to the incoming electron energies.

So far, we have not specified whether the vibrational states are bound or dissociative. In our study, the initial vibrational states are going to be bound, and in the final state, we consider both bound and dissociative ones. For the bound-bound transitions, Eq.~\eqref{eq:XS_in_terms_of_T} can be readily applied, replacing $i$ by $i\nu_i$ and $f$ by $f \nu_f$. For bound-dissociative transitions, a similar formula holds,
\begin{align}
    \dv{\sigma_{i \nu_i \rightarrow f \nu_E} (\varepsilon)}{\varepsilon^\prime}  = \frac{\pi}{k^2} \sum_{l m l^\prime m^\prime} |T_{i \nu_i l m \rightarrow f \nu_E l^\prime m^\prime} (\varepsilon,\varepsilon^\prime)|^2,
    %\tag{\ref{eq:XS_in_terms_of_T}$^\prime$}
    \label{eq:XS_in_terms_of_T_bc}
\end{align}
provided that the dissociative states $\nu_E$ are energy-normalized~\cite{stibbe_near-threshold_1998,scarlett_isotopic_2021}.

The dissociative character of the final states also hints at why, in Eq.~\eqref{eq:EB_AN}, we should use the correct outgoing electron energy instead of the incoming electron energy.
%The final dissociative states also motivate, that in Eq.~\eqref{eq:EB_AN}, we take the \say{correct} outgoing electron energy instead of the incoming electron one.
In our case, the fixed-nuclei incoming electron energy $\tilde{\varepsilon}$ differs from the correct one, $\varepsilon$, on the energy scale of the dissociation threshold of the initial electronic state. If we took the correct incoming electron energy, the fixed-nuclei outgoing energy would differ from the correct one on the scale of the dissociation energy. This discrepancy varies for different $\varepsilon^\prime$ with a fixed $\varepsilon$ and is severe when most of the energy is released as KER and the outgoing electron is slow. For the ICEC scenarios which correspond to electronic deexcitation, we have to use AN because EBAN introduces an unphysical threshold. %In the system studied in this paper, introduced below in Sec.~\ref{sec:system}, we therefore use EBAN for transitions X to B and A to B, and AN for transitions B to X and B to A.

With both the analytical model and the R-matrix approach in place, let us introduce the physical system we shall investigate.

\section{The system studied: $(\text{He}\text{Ne})^+$}
\label{sec:system}
In the previous sections, we introduced two approaches for calculating vibrationally-resolved ICEC cross sections.
Section~\ref{sec:theory} presented an analytical model extended to include nuclear dynamics, accounting for both energy and electron transfer mechanisms.
In Sec.~\ref{sec:methods_R-matrix}, we built upon the \textit{ab initio} fixed-nuclei R-matrix results and included the nuclear dynamics using the AN or EBAN approximation, referred to as the R-matrix.  
We apply both methods to the \ce{(HeNe)+} dimer as an example of a weakly bound system.

To contextualize the calculations, we begin with the structural and electronic properties of \ce{(HeNe)+}. 
Neon and helium have ionization potentials of \SI{21.56}{eV} and \SI{24.59}{eV}, respectively. 
Energy conservation implies that when \ce{He+} captures an electron, the electron affinity is sufficient to ionize neon. 
In contrast, electron capture by \ce{Ne+} requires additional kinetic energy.

Vibrational dynamics are described using Morse potentials for the three lowest electronic states, offering analytic solutions. 
Parameters from Ref.~\onlinecite{seong_hene_1999} are used and listed in Table~\ref{tab:params_orig}. 
Without loss of generality, we assume the dimer is aligned along the $z$-axis.
The ground state X asymptotically corresponds to $\mathrm{He}\mathrm{Ne}^+(2\mathrm{p_z^{-1}})$, with the charge localized on neon, also noted in the last column of Table~\ref{tab:params_orig}.
The first excited state A -- not to be confused with the electron acceptor -- corresponds to $\mathrm{He}\mathrm{Ne}^+(2\mathrm{p_{x/y}^{-1}})$, sharing the same asymptotic energy as X. 
The second excited state B, $\mathrm{He}^+(1\mathrm{s}^{-1})\mathrm{Ne}$, is the first with the charge on helium.

\begin{figure}[ht]
    \centering
    \includegraphics[width=0.4\linewidth, trim=10 10 10 0, clip]{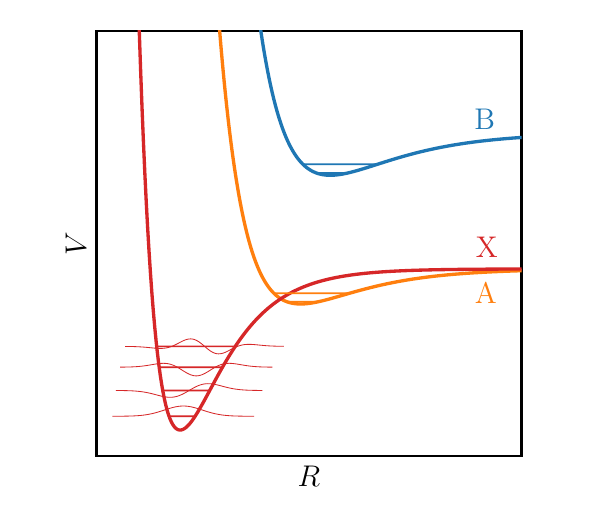}
    \caption{Potential energy surfaces X, A, and B corresponding to the three lowest electronic states of \ce{(HeNe)+}. For ease of comparison, surfaces A and B are multiplied by 4, and B is shifted down in energy. Figure taken from Ref.~\onlinecite{figures}.}
    \label{fig:potentials}
\end{figure}

\begin{table}[ht]
    \centering
    \caption{Parameters of the surfaces $\mathrm{X}$, $\mathrm{A}$, and $\mathrm{B}$ from Ref.~\onlinecite{seong_hene_1999}: equilibrium distance $R_\text{e}$, dissociation energy $D_\mathrm{e}$, harmonic vibrational frequency $\omega_\mathrm{e}$, and electronic energy $E$ at equilibrium. Last column denotes corresponding asymptotic ($R \rightarrow \infty$) states.}
    \begin{tabular}{ccccc@{\hspace{15pt}}c}
        \toprule
         state & $R_\text{e} [\unit{\angstrom}]$ & $D_\mathrm{e} [\unit{cm^{-1}}]$ & $\omega_\text{e} [\unit{cm^{-1}}]$ & $E [\unit{\atomicunit}]$ & asymptotics \\ 
         \midrule
         $\mathrm{X} \, ^2\Sigma^+$ & 1.43 & 5200 & 911 & -130.94349  & $\ce{He} + \ce{Ne+}(2\mathrm{p_z^{-1}})$  \\
         $\mathrm{A} \, ^2\Pi_{1/2}$& 2.42 &  283 & 152 & -130.92176  & $\ce{He} + \ce{Ne+}(2\mathrm{p_{x/y}^{-1}})$\\
         $\mathrm{B} \, ^2\Sigma^+$ & 2.66 &  343 & 152 & -130.80754  & $\ce{He+}(1\mathrm{s}^{-1}) + \ce{Ne}$ \\
         \bottomrule
    \end{tabular}
    \label{tab:params_orig}
\end{table}

This system provides an excellent platform for studying a wide range of vibrational effects, as it features potential energy surfaces with distinct characteristics. 
The excited states A and B are shallower and shifted toward larger internuclear distances compared to X (see Fig.~\ref{fig:potentials}), providing a diverse range of conditions found in different physical systems.

\section{Computational details}
\label{sec:computational_details}

In this section, we outline the computational framework used to model ICEC in the \ce{(HeNe)+} dimer. 
We first describe the numerical implementation of the analytical model, followed by the details of the R-matrix calculations.

%% THE MODEL
\subsection{The analytical energy and electron transfer model}
\label{methods:model}

To implement the model, we require (i) atomic photoionization cross sections for the energy transfer term in Eq.~\eqref{eq:en_transfer}, (ii) Gaussian widths for the electron transfer term in Eq.~\eqref{eq:el_transfer}, and (iii) vibrational states of the dimer.
The implementation~\cite{model-implementation} was carried out in Python using the \verb|mpmath| module~\cite{mpmath}, which supports confluent hypergeometric functions with complex arguments -- necessary for the dissociative Morse states adopted from Ref.~\onlinecite{matsumoto_1988}.

Photoionization cross sections for \ce{He} and \ce{Ne} were taken from Ref.~\onlinecite{photoionization_1976}. 
Since the three degenerate $2p$ orbitals in \ce{Ne} contribute equally under an isotropic field approximation \cite{sobelman_1973}, we divide the total cross section by three: $\sigma^{\text{PI}}_{2p\ell} = \sigma^\text{PI}/3$.
%This approximation is valid for transferred energies significantly lower than those of X-rays.

For the electron transfer contribution, we follow Ref.~\onlinecite{senk2024}, using Gaussian widths in Eq.~\eqref{eq:S_AD} based on covalent radii~\cite{covradii} scaled by $\alpha = 1.6$. 
Electron transfer efficiency depends on orbital orientation: in the ground electronic state (X), the $2p_z$ orbital of \ce{Ne+} aligns with the interatomic axis and facilitates the transfer. 
In the excited A state, the perpendicular $2p_{x/y}$ orbital results in negligible overlap with \ce{He}'s $s$-orbital, suppressing electron transfer. 
Consequently, we include electron transfer only in X to B and B to X transitions and restrict A to B and B to A transitions to the energy transfer mechanism.

We use the analytical bound and dissociative solutions of the Morse potential from Ref.~\onlinecite{matsumoto_1988}. 
Dissociative states are discretized within a finite box, $R \in (0,\SI{10}{\angstrom}]$. All but the highest bound vibrational state of each electronic state remain well confined within this range. 
We therefore exclude the highest initial bound state due to unreliable overlap with dissociative states in the finite box; this is justified at low temperatures where its population is negligible. 
%We exclude the highest bound states as initial states because their overlap with dissociative states is not well represented within the finite box, making the integration over $R$ unreliable. 
Final vibrational states are not restricted, as long as only bound initial states are considered -- excluding the highest initial state. 
%In this case, the spatial confinement of the initial wave function ensures that contributions outside the well of the initial Morse potential, and thus outside the box, remain negligible.
Energy-normalized cross sections are obtained by multiplying box-normalized values by the density of states, $\rho(E_i) = 2 / |E_{i+1} - E_{i-1}|$.

Unless stated otherwise, we assume the dimer starts in the vibrational ground state of the initial electronic state and sum over all final vibrational states.
For fixed-nuclei calculations, the ICEC cross section is evaluated at the equilibrium distance $R_\text{e}$ of the initial electronic state.

%% THE FIXED-NUCLEI R-MATRIX
\subsection{The R-Matrix approach}

We employed the fixed-nuclei R-matrix method as implemented in the \texttt{UKRmol+} suite~\cite{masin_ukrmol_2020}. Molecular orbitals for the scattering calculations were obtained using the \texttt{MOLPRO}~\cite{werner_molpro_2012, werner_molpro_2020, werner_molpro_nodate} implementation of the Hartree-Fock method applied to the $\ce{HeNe}$ system, with the cc-pVDZ basis set. Continuum orbitals were constructed using a radial GTO basis optimized for an R-matrix radius of $a = \qty{13}{\bohr}$, as provided in the \texttt{UKRmol-scripts} release~\cite{houfek_ukrmol-scripts_2024}, and included partial waves up to angular momentum $l = 6$.
%We used the fixed-nuclei R-matrix implementation of the \texttt{UKRmol+} suite~\cite{masin_ukrmol_2020}.
%First, we calculated the molecular orbitals entering the scattering model.
%We used the cc-pVDZ basis. The molecular orbitals were optimized using \texttt{MOLPRO}~\cite{werner_molpro_2012,werner_molpro_2020,werner_molpro_nodate} for the $\ce{HeNe}$ system using the Hartree-Fock method.
%For the continuum orbitals, we used the radial GTO basis optimized for the R-matrix radius $a=\qty{13}{\bohr}$ included in the \texttt{UKRmol-scripts}~\cite{houfek_ukrmol-scripts_2024} release, with angular momentum up to $l=6$.

In the scattering model, we used the $1 \text{s}$ orbital of \ce{Ne} as frozen and the active space comprised the $2\text{s}$, $2\text{p}$ and $3\text{p}$ orbitals of \ce{Ne} and the $1\text{s}$ and $2\text{s}$ orbitals of \ce{He}. We included four target states into the scattering calculations, as only those are energetically accessible in the investigated region of scattering energies. The R-matrix was propagated to $r_\text{af}=\qty{80}{\bohr}$.

%% INCLUDING VIBRATIONAL MOTION INTO THE R-MATRIX

We performed the fixed-nuclei R-matrix calculations for interatomic distances ranging from $R=\qty{1.5}{\bohr}$ to $R=\qty{9.5}{\bohr}$ with a step of $\Delta R = \qty{0.5}{\bohr}$. To evaluate the dissociative states, we use the sine discrete variable representation (DVR)~\cite{colbert_dvr_1992} in the confinement box $R\in (0,10\,\text{\r{A}}]$ with $3000$ gridpoints.
%The dissociative states are renormalized into energy normalization by multiplying them by the square root of the numerically-evaluated density of DVR states.

We use the AN approximation for the transitions B to X and B to A and the EBAN approximation for the transitions X to B and A to B.
The $R$-integral in Eq.~\eqref{eq:AN} and Eq.~\eqref{eq:EB_AN} for AN and EBAN, respectively, is facilitated using a composite 10-point Gauss-Legendre quadrature~\cite{stoer_introduction_1993,abramowitz_handbook_1964}. Because the dissociative states in the energy range studied can oscillate with high frequencies, it is necessary to interpolate the fixed-nuclei T-matrices, because performing the \textit{ab initio} calculations on such a dense grid would be excessively computationally demanding. In the region of energies we investigate, there are no resonances, and the T-matrix elements are smooth both in energy and in $R$. The cubic spline interpolation~\cite{burden_numerical_2016} is used. The $R$-quadrature is carried out over the extent of the DVR box. For obtaining the T-matrix values outside the range of the fixed-nuclei calculations, we used linear extrapolation for $R < \qty{1.5}{\bohr}$ and $ 1 / R^3$ extrapolation for $R > \qty{9.5}{\bohr}$ motivated by the form of the asymptotic approximation~\cite{gokhberg2009} to fixed-nuclei ICEC cross sections that is valid in this range. The implementation is available at Ref.~\onlinecite{rmatrix-implementation}.

With the computational framework in place, we proceed to analyze the ICEC cross sections for the \ce{(HeNe)+} dimer.

\section{Results and discussion}
\label{sec:results}

In this section, we present the computed ICEC cross sections for the \ce{(HeNe)+} dimer, examining how vibrational motion affects the efficiency of electron capture. 
We compare results from the analytical model\cite{model-results} and the R-matrix approach\cite{rmatrix-results}, highlighting key trends and deviations.

We begin by discussing the energy transfer and electron transfer contributions to ICEC in Sec.~\ref{sec:energy_tf_vs_electron_tf}. In Sec.~\ref{sec:total_XS}, we investigate the total cross section for ICEC from the ground vibrational state and compare with the transitions from the excited vibrational states. Then, we present the temperature-dependent ICEC cross section in Sec.~\ref{sec:temperature_dependence}. Last but not least, we investigate the spectrum of the outgoing electron for a fixed incoming electron energy in Sec.~\ref{sec:spectrum}. 

\subsection{Energy vs. electron transfer for B to X and X to B}
\label{sec:energy_tf_vs_electron_tf}

% discuss one first, then tell what is different in the other case and wether it is to be expected (and why) or surprising $\to$ what do we learn?}
% Then you don't need the lengthy abstract list of which line is what, but you can discuss science as you introduce the different components.}

The analytical model, introduced in Sec.~\ref{sec:theory}, includes both energy and electron transfer contributions to ICEC. 
While energy transfer applies to all transitions, electron transfer is only relevant when the state X is involved (see Sec.~\ref{methods:model}). 
We therefore focus on the B to X and X to B transitions in this chapter.
%We therefore want to have a closer look at the B to X and X to B transitions.

\begin{figure}[ht]
\centering
    \begin{subfigure}{0.49\linewidth}
        \centering
        \includegraphics[width=\linewidth]{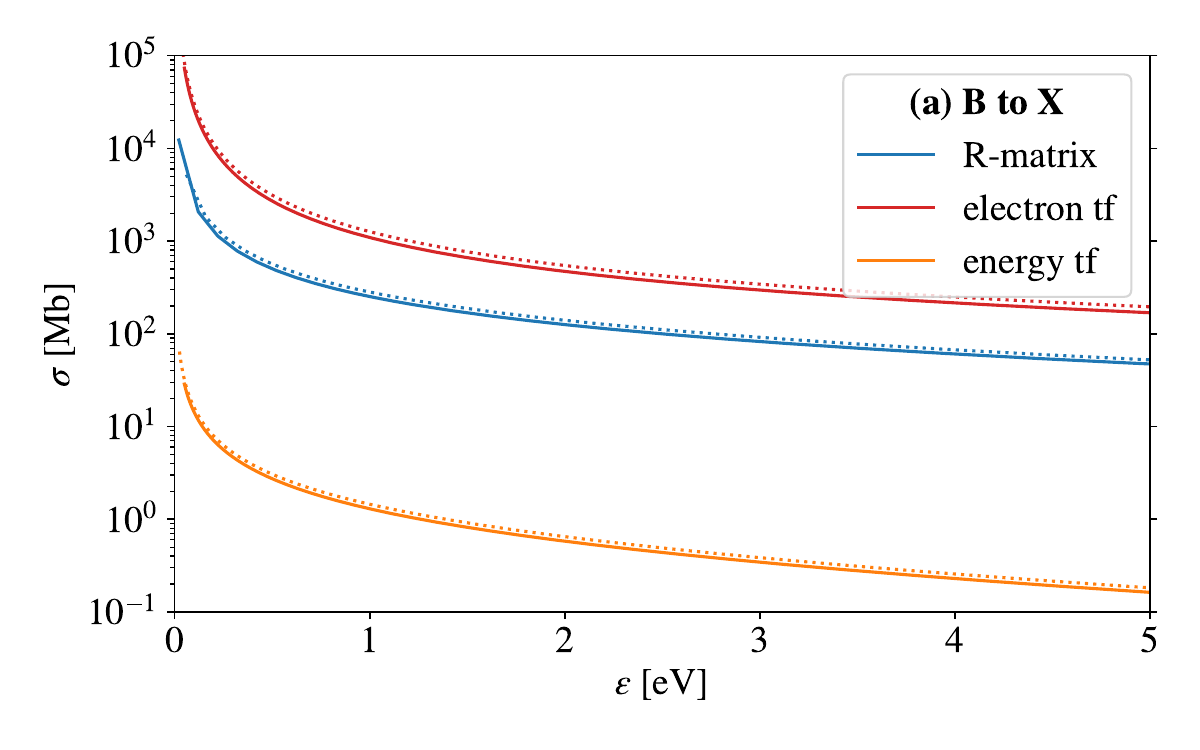}
    \end{subfigure}
    \begin{subfigure}{0.49\linewidth}
        \centering
        \includegraphics[width=\linewidth]{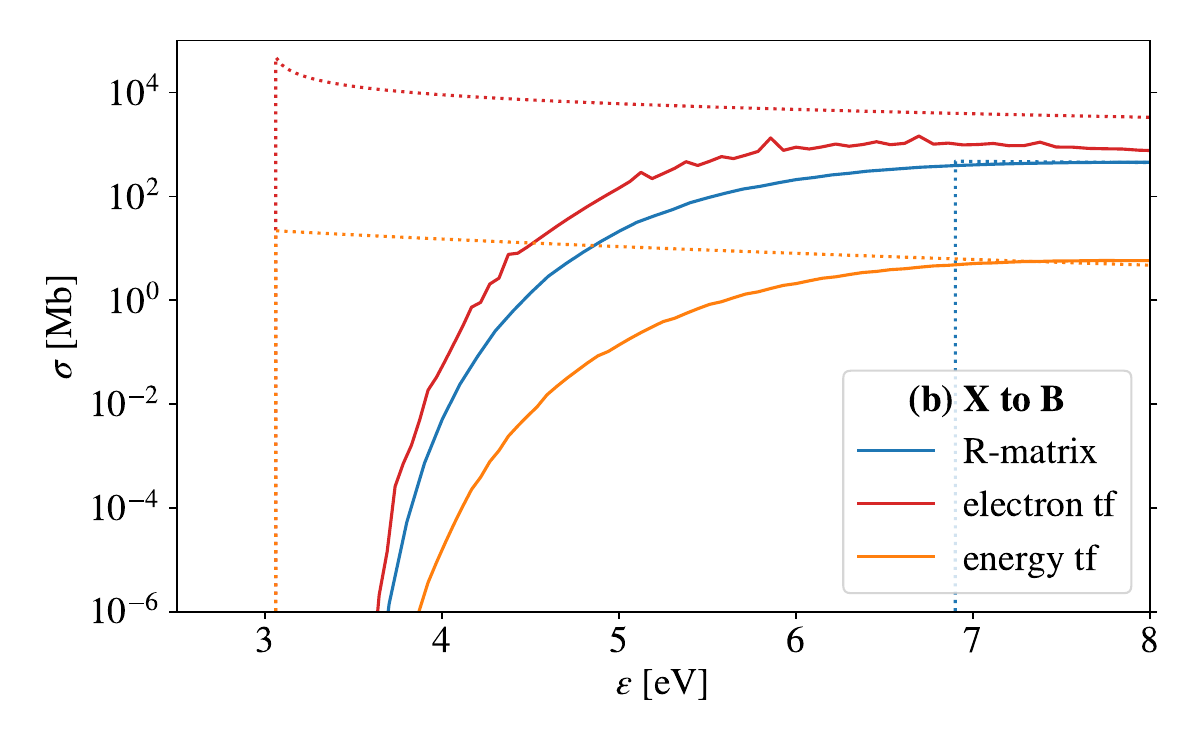}
    \end{subfigure}
    \caption{
    ICEC cross section vs. incoming electron energy $\varepsilon$ for $\nu_i=0$ from energy transfer (orange), electron transfer (red), and R-matrix (blue). Solid lines include nuclear dynamics; dotted lines are fixed-nuclei results. The fixed-nuclei model uses asymptotic energies while R-matrix considers a vertical transition at $R_e$, resulting in different thresholds for X to B. Figures taken from Ref.~\onlinecite{figures}}
    \label{fig:asympt-overlap}
\end{figure}

Figure~\ref{fig:asympt-overlap} shows the corresponding ICEC cross sections as a function of the incoming electron energy. 
The electron transfer contribution (red) dominates over energy transfer (orange) by three orders of magnitude, making it the primary mechanism for these transitions. 
This dominance stems from the electron transfer being strongly favored at short interatomic distances around $R_e$~\cite{senk2024}.
%Since the dimer starts in its ground vibrational state, the atoms remain near equilibrium separation $R_\text{e}$, where electron transfer is particularly efficient due to its favorable scaling with $R$ \cite{senk2024}. 
Despite their magnitude difference, both contributions show similar trends with increasing kinetic energy $\varepsilon$.

We compare the analytical model to the R-matrix results (blue), which include both energy and electron transfer contributions. 
%We compare the analytical model to the R-matrix results, which encompasses both the energy and electron transfer contributions.
While all curves exhibit the same overall trend, the analytical model is roughly a factor of three larger than the R-matrix results. 
%All curves show the same trend, however, the sum of the two contributions from the analytical model is approximately a factor of three larger than the R-matrix results.
This discrepancy stems from the electron transfer term, which is exaggerated due to parameter choices in the Gaussian description of the electronic distributions for the \ce{(HeNe)+} system.~\cite{senk2024}
%This is caused by the electron transfer contribution. Due to the choice of parameters for the descriptions of the electron distributions by the Gaussians, this contribution is overestimated for our system (HeNe$^+$). \cite{senk2024}

%Among the two components of the model, the electron transfer cross section shows better agreement with the R-matrix data (blue), staying within one order of magnitude, while the energy transfer significantly underestimates the cross section by about two orders.  
%The slight overestimation by electron transfer stems from the parameterization in Ref.~\onlinecite{senk2024}, where Gaussian widths were fitted linearly across systems. 
%In the case of \ce{(HeNe)+}, this results in exaggerated orbital overlap and consequently inflated cross sections.

From here on, we include both energy and electron transfer contributions in the model for the X to B and B to X transitions. 
We now turn to comparing the total cross section with vibrationally resolved and fixed-nuclei results.

\subsection{Total cross section}
\label{sec:total_XS}

\begin{figure}[!ht]
    \centering
    \begin{subfigure}{0.49\linewidth}
        \includegraphics[width=\linewidth, page=1]{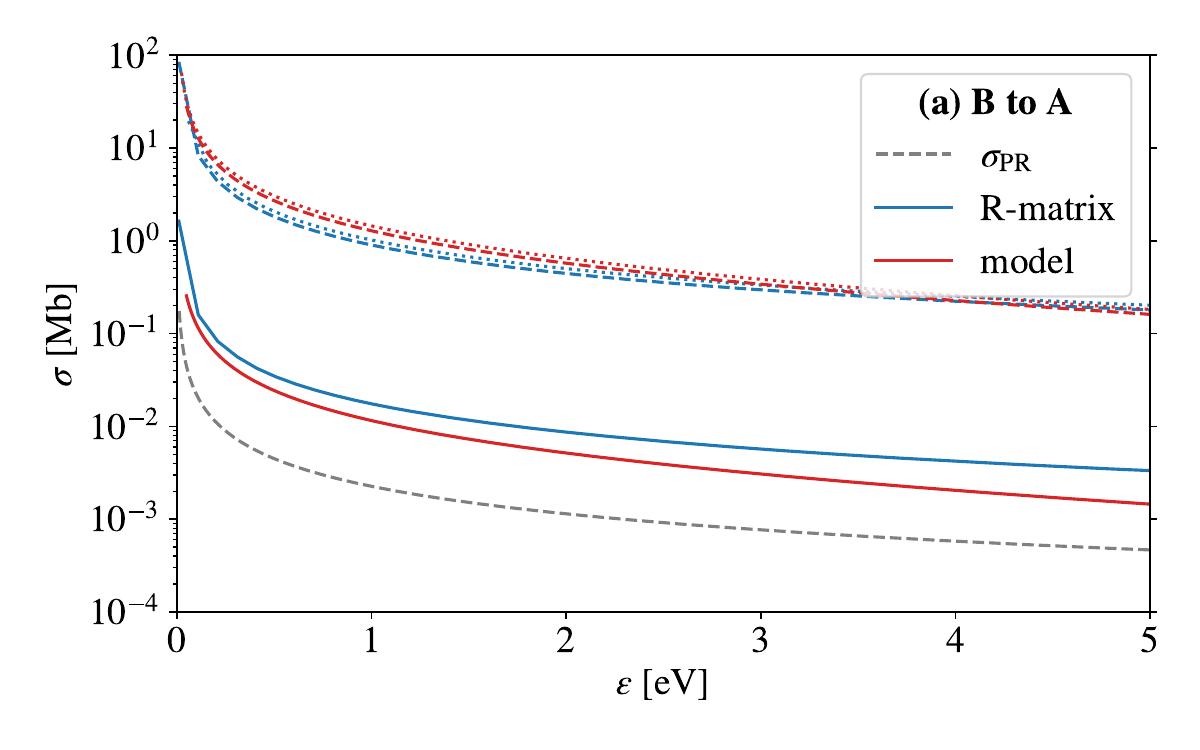}
    \end{subfigure}
    \begin{subfigure}{0.49\linewidth}
        \includegraphics[width=\linewidth, page=1]{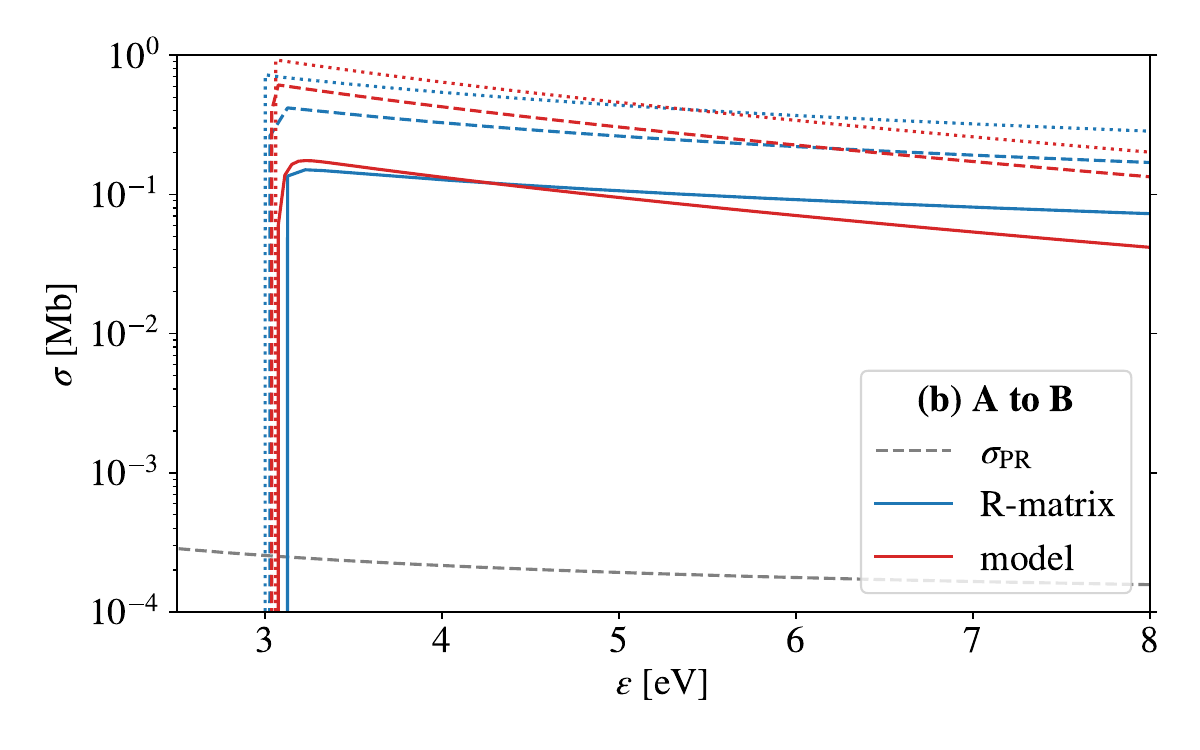}
    \end{subfigure}
    \begin{subfigure}{0.49\linewidth}
        \includegraphics[width=\linewidth, page=1]{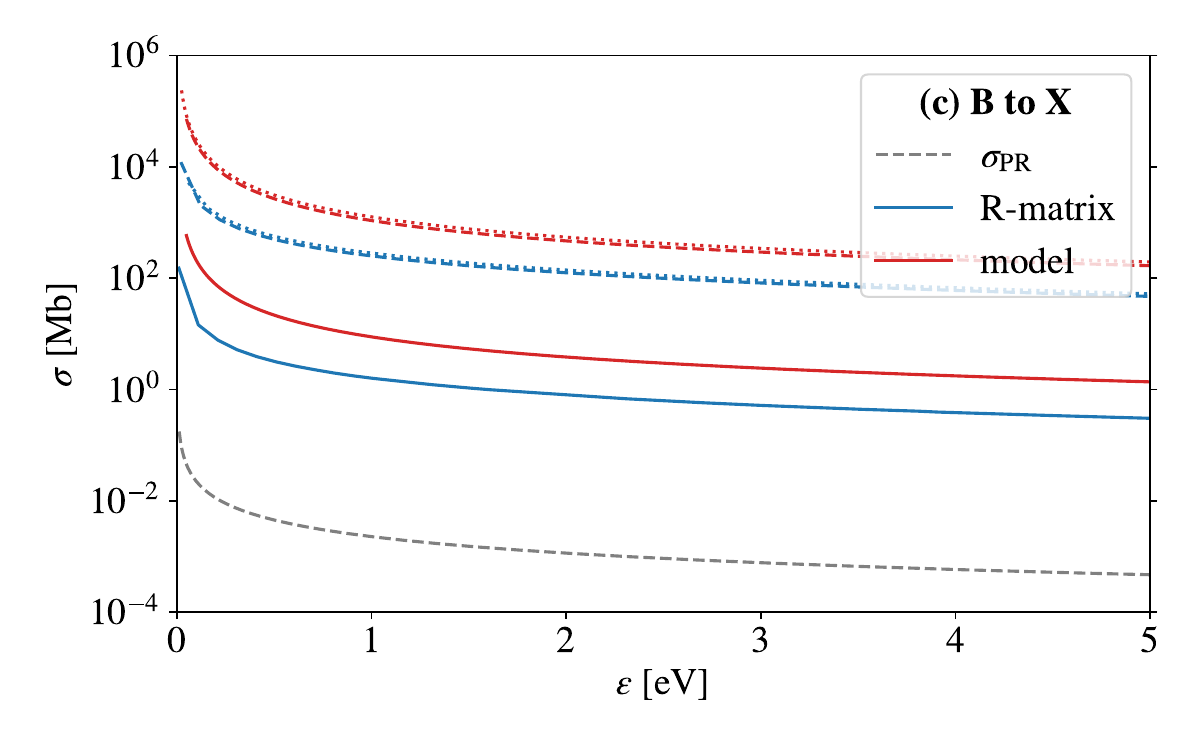}
    \end{subfigure}
    \begin{subfigure}{0.49\linewidth}
        \includegraphics[width=\linewidth, page=1]{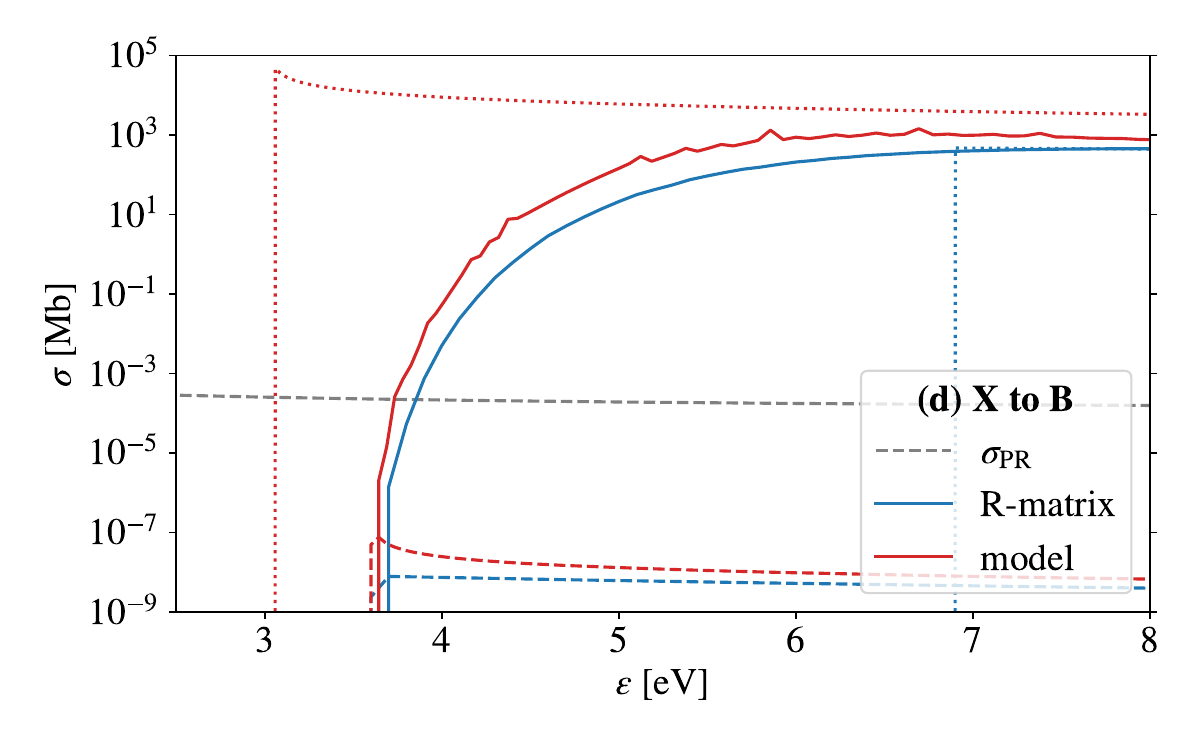}
    \end{subfigure}
    \caption{
    ICEC cross section vs. incoming electron energy $\varepsilon$ for $\nu_i = 0$. Left: electron captured by \ce{He+}; right: captured by \ce{Ne+}. R-matrix (blue) and model (red) results: solid for bound-dissociative, dashed for bound-bound transitions. Dotted lines show fixed-nuclei cross sections; dashed grey lines represent photorecombination cross sections on the capturing atom. Figures taken from Ref.~\onlinecite{figures}}
    \label{fig:tot_xs}
\end{figure}

Figure~\ref{fig:tot_xs} presents ICEC cross sections for all electronically distinct processes, comparing the analytical model, the R-matrix, and fixed-nuclei results. 
Across all cases, the model agrees with the R-matrix approach within an order of magnitude, consistent with the comparisons discussed in Sec.~\ref{sec:energy_tf_vs_electron_tf}.

Threshold energies differ between channels. 
When \ce{He+} captures the electron (left panels), the process can occur at zero incoming electron energy due to the high electron affinity of \ce{He+}. 
For the reverse process (right panels), the incoming electron has to supply an energy of $3-\qty{4}{\electronvolt}$ to make ICEC possible.

The relative importance of bound-bound (b-b) and bound-dissociative (b-d) transitions varies by channel. 
For B to X and B to A (left panels), b-b transitions dominate by about two orders of magnitude, driven by large overlaps between the ground vibrational state of B and bound states of X or A. 
%In X, large contribution is associated with higher excited vibrational states.
For A to B, b-b transitions still dominate but only marginally.

In contrast, the X to B process is dominated by b-d transitions -- by up to ten orders of magnitude. 
The suppression of the b-b transitions arises from minimal overlap between the ground state of X and the bound states of B, due to the shift in equilibrium distance $R$, see Fig.~\ref{fig:potentials}. 
The same holds even for low-energy dissociative states but the overlap grows with increasing KER (the energy of the dissociative state), as the turning point moves to smaller $R$. 
%At low energies, the overlap with the available dissociative states increases gradually as the turning point moves to smaller $R$. 
%With increasing KER (the energy of the dissociative state), the turning point of the dissociative states reaches to lower $R$, increasing the overlap with the ground vibrational state of X. 
%This is the reason behind the gradual increase of the cross sections of the bound-dissociative transitions above the threshold. 
%The more energy is available after recombination, the more dissociative states can be reached. 
With larger kinetic energy $\epsilon$, more dissociative states are accessible, enhancing the total cross section.
Once the relevant dissociative states are fully accessible, the cross section saturates and eventually decreases due to the kinetic energy dependence.
%Once all dissociative states of B with significant overlap with the ground vibrational state of X are available, the increase is saturated, and increasing the energy does not increase the cross section further. 
%Instead, it starts to decrease because of the underlying energy dependence in the fixed-nuclei cross sections, see also Fig.~\ref{fig:asympt-overlap}.

Comparison with fixed-nuclei results calculated at the equilibrium distance of the initial electronic state (dotted lines) reveals the impact of nuclear dynamics. 
%Let us now discuss how including the vibrational dynamics affects the cross sections by comparing them with the fixed-nuclei results at $R_\text{e}$ represented by the dotted lines. 
For transitions B to A, B to X, and A to B, the fixed-nuclei cross sections are comparable to the dominant b-b or b-d contribution. 
For A to B, the agreement in magnitude is within a factor of two across the full energy range. %The threshold behaviors are also similar.
%In all of the cases, this comparison is valid in the whole plotted range of incoming electron energies, as the threshold behavior of the vibrationally resolved cross section and the fixed-nuclei one is the same. 

The X to B transition, however, shows a significant discrepancy in threshold between the fixed-nuclei and vibrationally resolved cross sections. 
This stems from different energy conservation assumptions: the fixed-nuclei model (dotted red) uses asymptotic energy conservation, $\epsilon'= \epsilon + \text{IP}_{\text{A}^-} - \text{IP}_\text{D}$\cite{gokhberg2009}, while the fixed-nuclei R-matrix  (dotted blue) uses vertical transitions at $R_\text{e}$, see Eq.~\eqref{eq:AN_fixed_nuclei_energy_conservation}.
%The R-matrix threshold at $R_\text{e}$ is $\approx \qty{7}{\electronvolt}$. 
Both the model and R-matrix approaches for including nuclear dynamics instead conserve the vibronic energy, Eq.~\eqref{eq:energy_conservation_A}, yielding a more gradual onset of the cross section.
%For X to B, the inclusion of nuclear dynamics also affects the threshold behavior, causing a gradual onset for the bound-dissociative transitions, as we already discussed above. 
%For incoming electron energies higher than $\approx \qty{7}{\electronvolt}$, the fixed-nuclei results agree with the vibrational ones in shape and magnitude as for the other processes.

%Moreover, in both panels of Fig.~\ref{fig:all-vi}, we see that including nuclear dynamics decreases the cross sections in comparison with the fixed-nuclei prediction at the equilibrium of the initial state.

In all cases, the dominant ICEC cross section -- whether b-b or b-d -- surpasses the photorecombination cross section of the capturing ion by at least three orders of magnitude. 
For the X to B transition, the enhancement reaches up to six orders of magnitude, highlighting the efficiency of ICEC and its significant role in environments where low-energy electrons and weakly bound systems are present.
%potential observability in future experiments.

\subsection{Temperature dependence}
\label{sec:temperature_dependence}

To model ICEC realistically, the thermal population of initial vibrational states must be considered. 
Unlike fixed-nuclei methods, our approach includes vibrational averaging via a Boltzmann distribution, introducing temperature dependence into the cross sections, as outlined in Sec.~\ref{sec:theory_considerations}. 

\begin{figure}[!ht]
    \centering
    \begin{subfigure}{0.49\linewidth}
        \includegraphics[width=\linewidth]{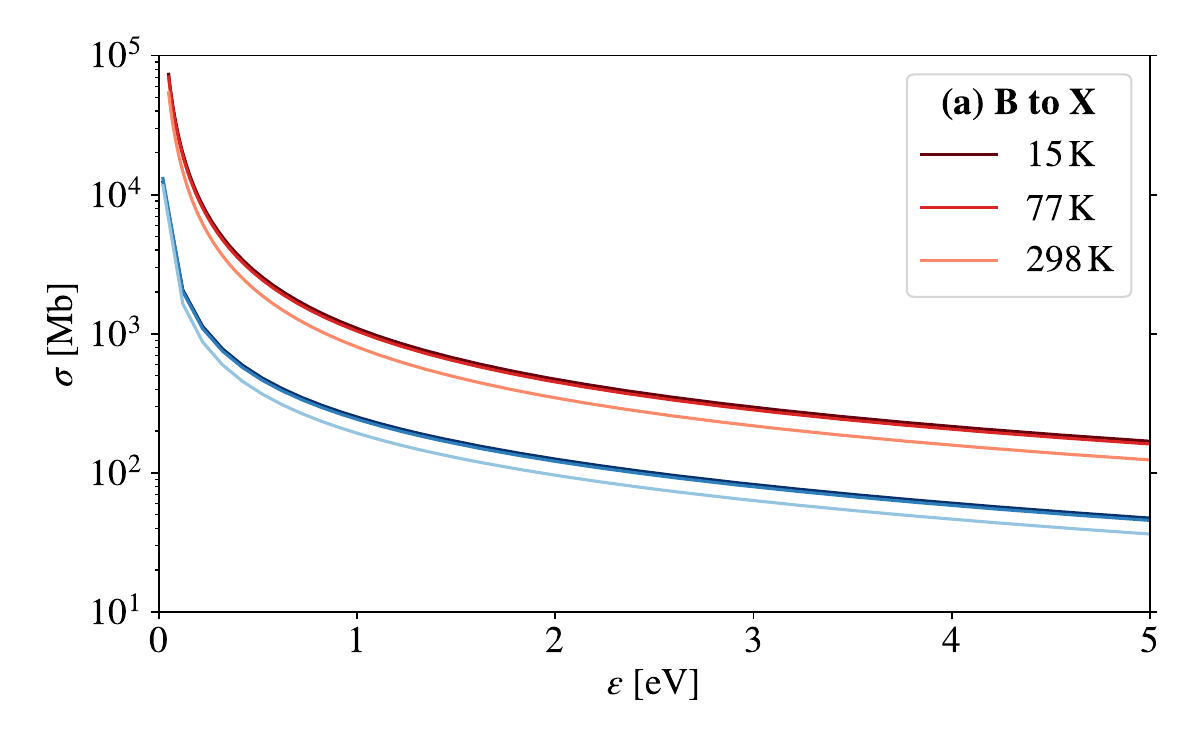}
    \end{subfigure}
    \begin{subfigure}{0.49\linewidth}
        \includegraphics[width=\linewidth]{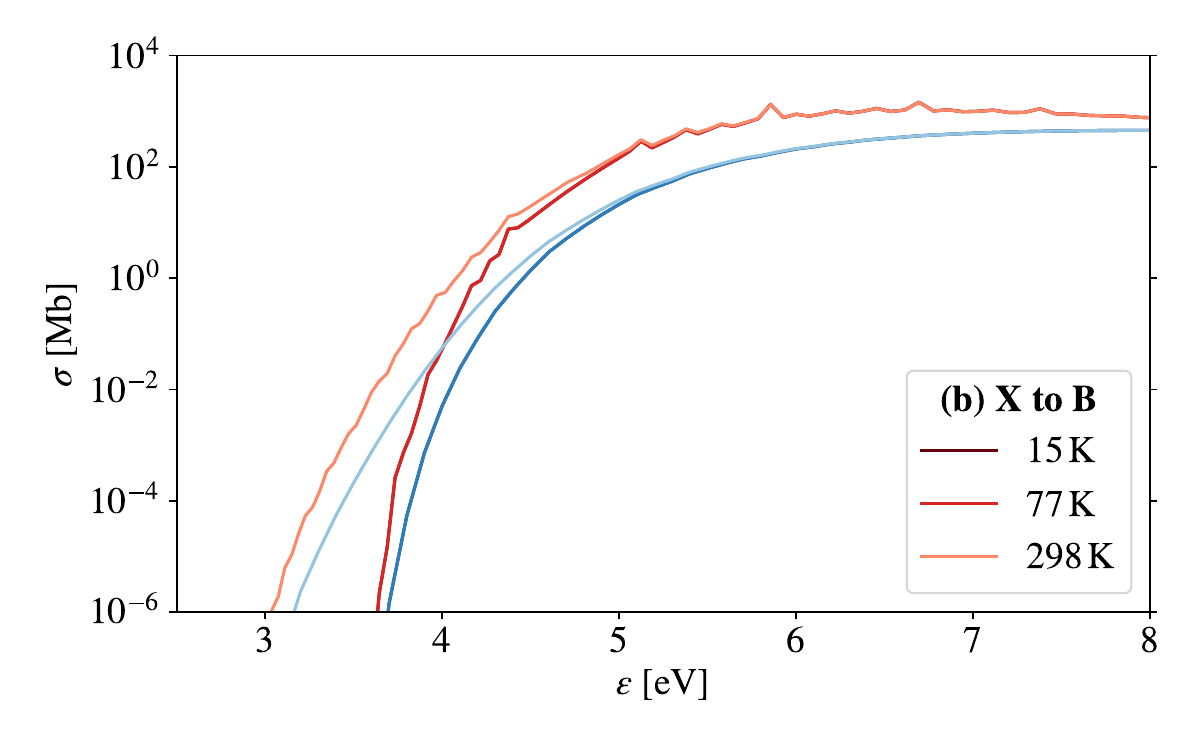}
    \end{subfigure}
    \caption{
    Temperature-dependent ICEC cross section vs. incoming electron energy $\varepsilon$. Shades of blue: the R-matrix; shades of red: the model. Lighter shades correspond to higher temperatures. The curves for $T=\SI{15}{K}$ and $\SI{77}{K}$ coincide in the right panel. Figures taken from Ref.~\onlinecite{figures}}
    \label{fig:boltzmann}
\end{figure}

Figure~\ref{fig:boltzmann} illustrates how the ICEC cross section for \ce{(HeNe)+} varies with temperature. For the B to X, A to B, and B to A transitions, increasing the temperature to $T = \SI{298}{K}$ results in a reduction of the total cross section. For instance, in the B to X transition (panel~\ref{fig:boltzmann}a), the asymptotic model predicts a decrease of up to a factor of 1.4 when comparing $\SI{298}{K}$ (orange) to $\SI{15}{K}$ (red). This trend is also observed in the R-matrix calculations (blue).
The decrease stems from significant population of excited vibrational states at higher temperatures, and for these transitions, excited initial states yield lower cross sections.  
A detailed analysis of this vibrational-state dependence is provided in the Supplementary Material.

In contrast, the X to B transition (panel~\ref{fig:boltzmann}b) exhibits the opposite behavior. At low incoming electron energies, the cross section increases with temperature. 
This enhancement is due to the strong contribution from higher initial vibrational states, which, in this case, produce substantially larger cross sections in the low-energy regime due to lowering the threshold and enhancing the overlap between initial and final vibrational wave functions. More details can be found in the Supplementary Material.
%This behavior is also explained in the Supplementary Material. 
At higher electron energies, where the dependence on the initial vibrational state becomes weaker, the temperature effect diminishes and the cross section remains nearly unchanged.

Moreover, for X to B, we observe a lowering of the ICEC threshold with increasing temperature. This is a direct consequence of the reduced threshold energies associated with vibrationally excited initial states, which begin to dominate the vibrational average at higher temperatures.

%Overall, incorporating nuclear motion and temperature-dependent vibrational averaging leads to important modifications of the ICEC cross sections. These include changes in threshold behavior, energy dependence, and absolute magnitude. While nuclear dynamics can moderately suppress the cross section in some transitions, ICEC remains several orders of magnitude more efficient than photorecombination across all examined conditions.

\subsection{Spectrum}
\label{sec:spectrum}

Nuclear dynamics significantly impacts the spectrum of the outgoing electron in ICEC. At fixed incident electron energy, the spectrum features discrete peaks from bound-bound transitions and a continuum from bound-dissociative transitions.

\begin{figure}[!ht]
    \centering
    \begin{subfigure}{0.49\linewidth}
        \includegraphics[width=\linewidth]{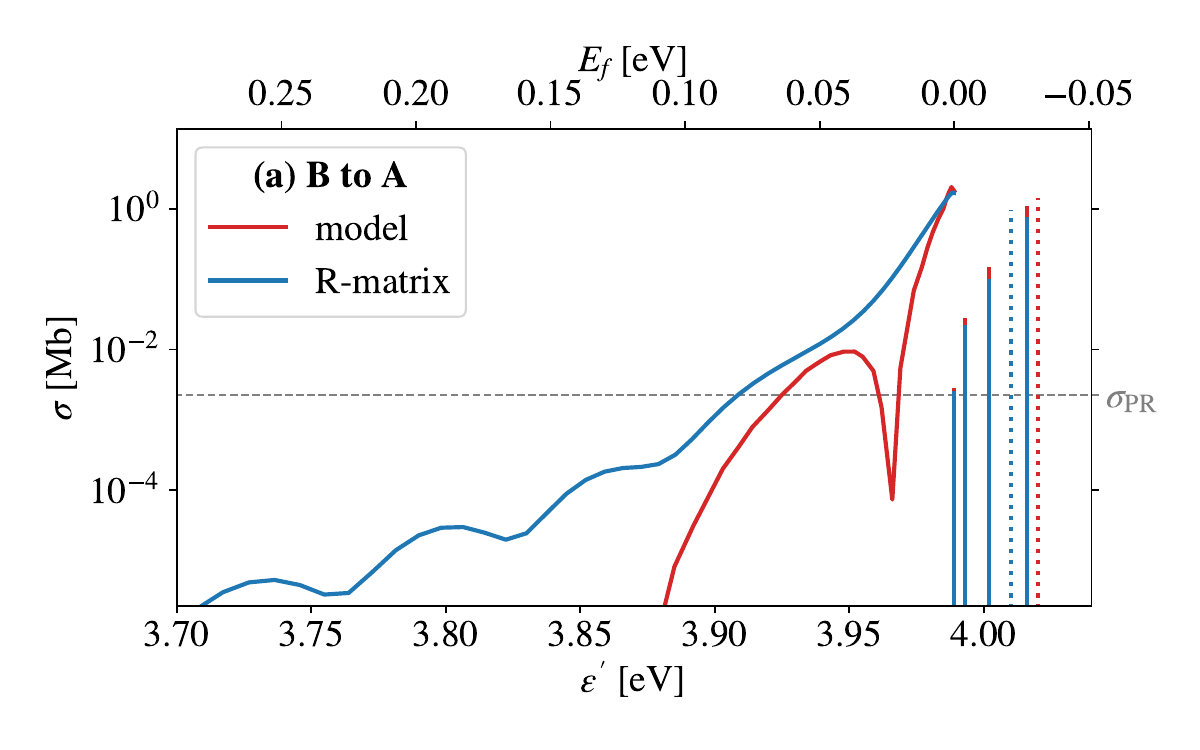}
    \end{subfigure}
    \begin{subfigure}{0.49\linewidth}
        \includegraphics[width=\linewidth]{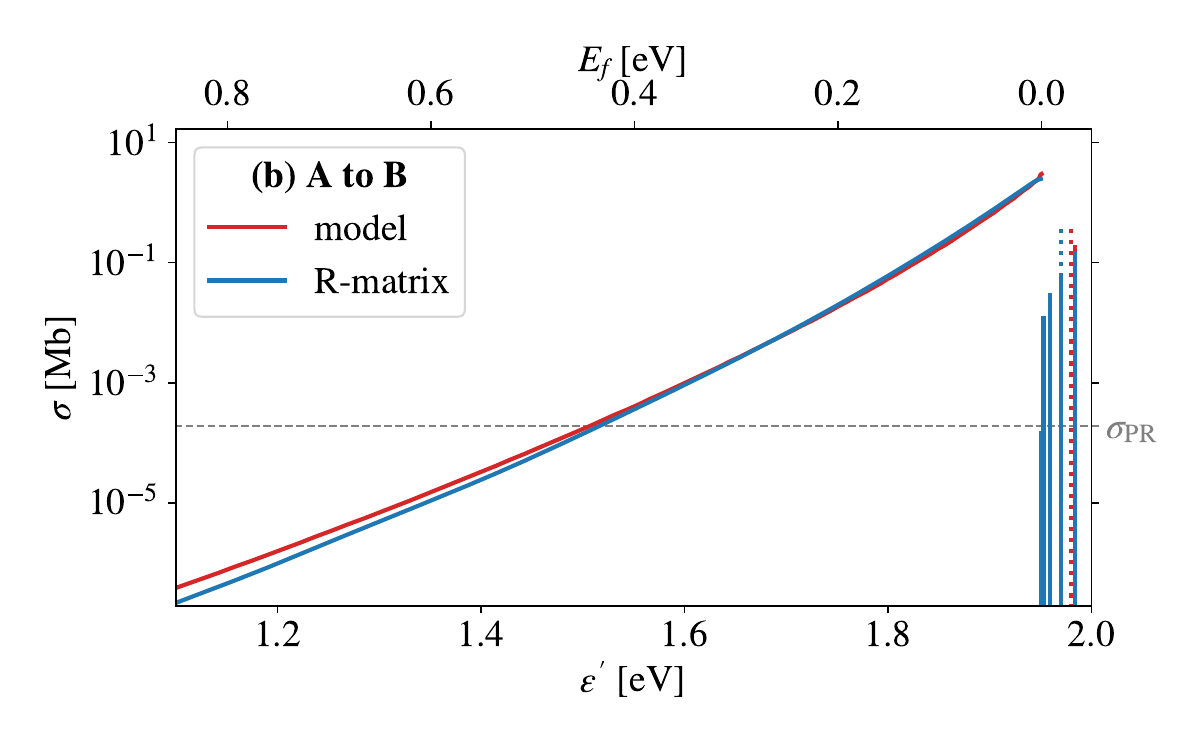}
    \end{subfigure}
    \begin{subfigure}{0.49\linewidth}
        \includegraphics[width=\linewidth]{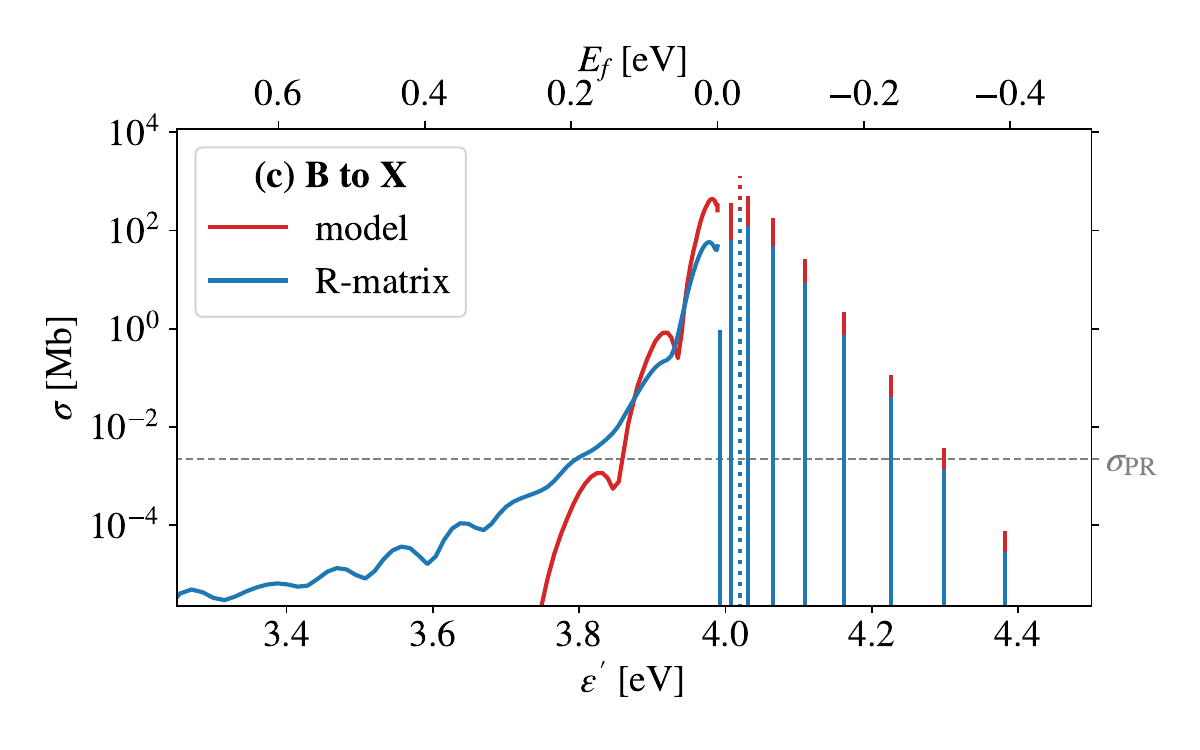}
    \end{subfigure}
    \begin{subfigure}{0.49\linewidth}
        \includegraphics[width=\linewidth]{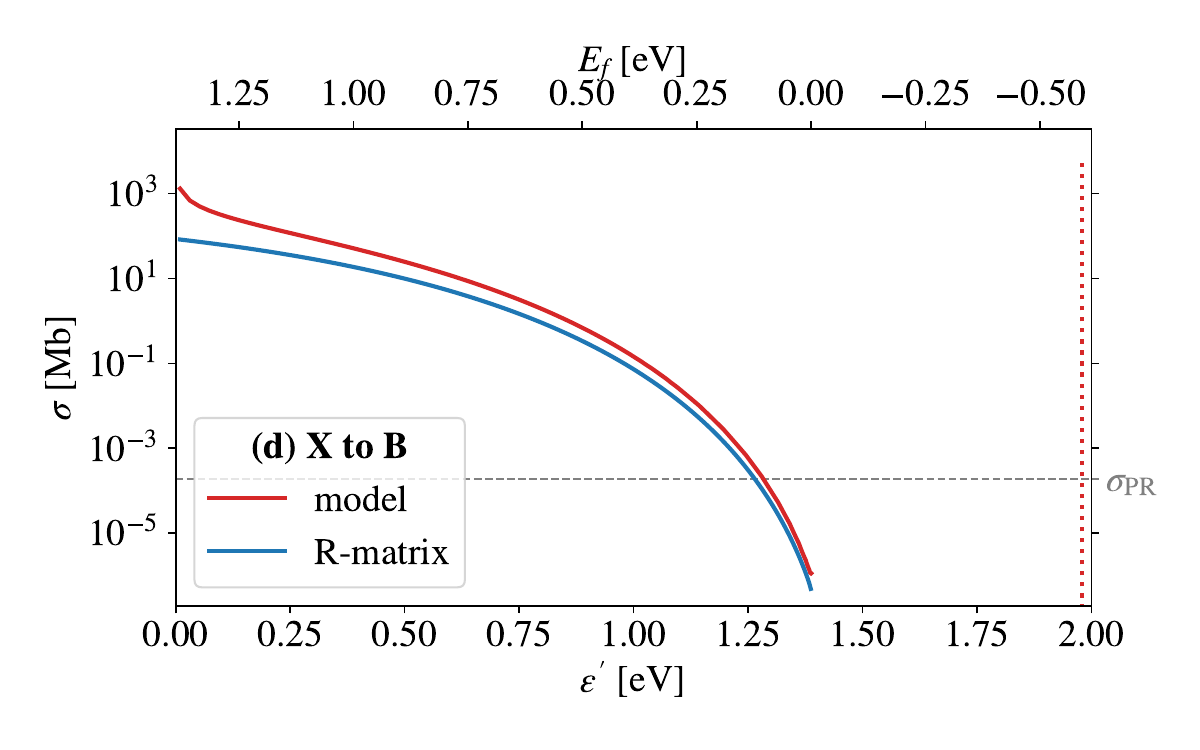}
    \end{subfigure}
    \caption{
    ICEC cross section vs. outgoing electron energy $\varepsilon'$ for $\nu_i = 0$. Each panel shows a different electronic ICEC process: left for \ce{He+} ($\varepsilon = \qty{5}{\electronvolt}$), right for \ce{Ne+} ($\varepsilon = \qty{1}{\electronvolt}$). Sticks denote bound-bound transitions; solid lines represent bound-dissociative differential cross sections in $\unit{\mega \barn \per \electronvolt}$, see Eqs.~\eqref{eq:sum_over_v} and~\eqref{eq:XS_in_terms_of_T_bc}. R-matrix (blue), model (red), photorecombination reference (dashed grey), and fixed-nuclei data (dotted sticks) are included. Figures taken from Ref.~\onlinecite{figures}}
    \label{fig:spectrum_v0}
\end{figure}

Figure~\ref{fig:spectrum_v0} presents the outgoing electron energy spectra for all ICEC processes considered, assuming initial ground vibrational states. 
We use $\varepsilon = \qty{1}{\electronvolt}$ (\ce{He+} as acceptor) and $\varepsilon = \qty{5}{\electronvolt}$ (\ce{Ne+} as acceptor). 
Sticks denote bound-bound transitions, with the rightmost corresponding to the ground vibrational final state; the continuous lines represent bound-dissociative contributions. Spectra are truncated below $10^{-3} \times \sigma_\text{PR}$ due to experimental irrelevance.

For A to B, B to A, and B to X, bound-dissociative cross sections decrease with increasing KER, often with modulations. This behavior persists for higher initial vibrational states (see Supplementary Material). For X to B, by contrast, the bound-dissociative cross section increases rapidly with KER. Higher $\nu_i$ introduce modulations, with an increasing number of minima.

\begin{figure}[!ht]
    \centering
    \begin{subfigure}{0.5\linewidth}
        \includegraphics[width=\linewidth]{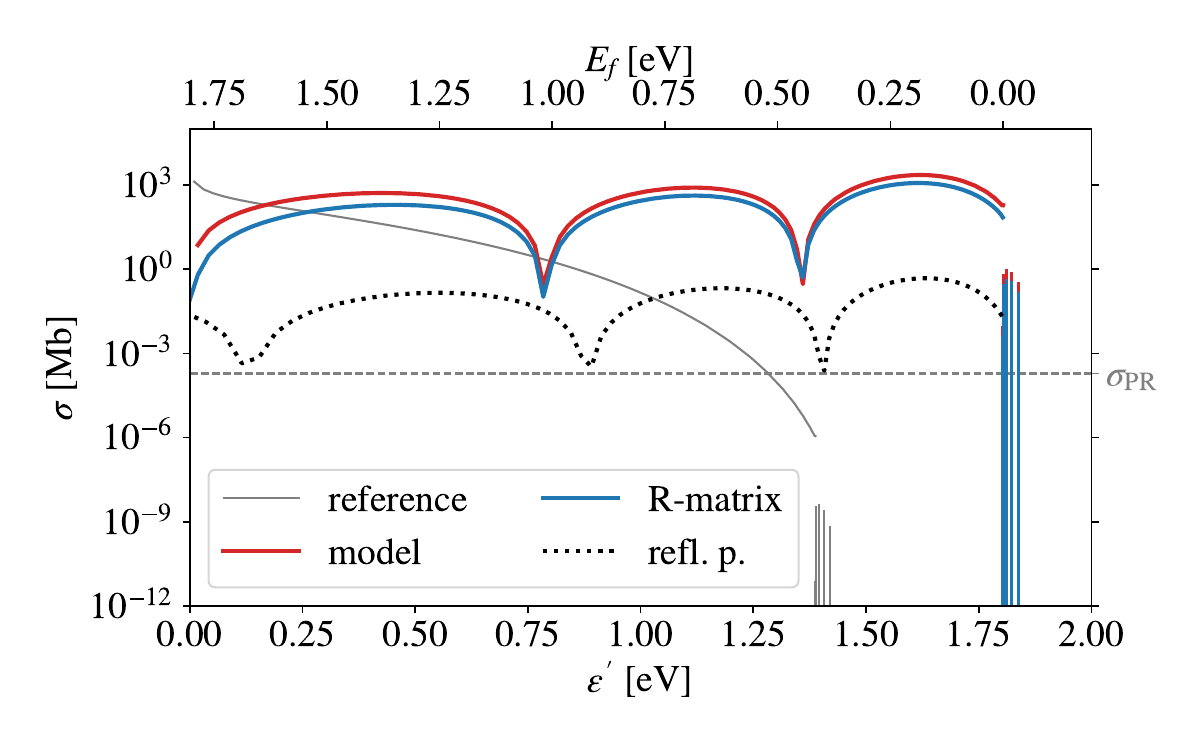}
    \end{subfigure}
    \caption{ICEC cross section vs. outgoing electron energy $\varepsilon^\prime$ for X to B. $\ce{Ne+}$ captures the incident electron with a kinetic energy of $\varepsilon = \qty{5}{\electronvolt}$. The initial vibrational state is $\nu_i = 5$. 
    The dotted line represents the reflection principle prediction of the cross section. The reference (grey line) shows the model data for $\nu_i=0$ (i.e. from panel~\ref{fig:spectrum_v0}d). The blue and red solid lines and sticks are as in Fig.~\ref{fig:spectrum_v0}. Figure taken from Ref.~\onlinecite{figures}}
    \label{fig:spectrum_X_B_vi5}
\end{figure}

Figure~\ref{fig:spectrum_X_B_vi5} compares X to B spectra for $\nu_i = 0$ and $\nu_i = 5$. Three key differences arise: (i) The entire spectrum for $\nu_i = 5$ is shifted by $\sim \qty{0.4}{\electronvolt}$ due to the initial vibrational energy difference; (ii) bound-bound cross sections increase by eight orders of magnitude for $\nu_i = 5$, due to enhanced overlap with final bound states at larger $R$; (iii) the dissociative spectrum for $\nu_i = 5$ shows multiple peaks and valleys, in contrast to the monotonic shape for $\nu_i = 0$.

This structure can be qualitatively explained using the reflection principle~\cite{gislason_series_1973}. By projecting the initial vibrational state's density, weighted by the electron transfer interaction $\propto \exp(-c R^2)/R^2$, see Eqs.~\eqref{eq:S_AD}, \eqref{eq:el_transfer}, we reproduce the overall shape and nodal structure of the cross section. However, the predicted peak positions are systematically shifted to higher $\varepsilon^\prime$, likely due to the approximation of the final-state potential by its linear component at the turning point within the reflection principle approximation.

Finally, we compare the analytical model and the R-matrix results. 
Generally, the model reproduces R-matrix cross sections within an order of magnitude. The largest discrepancies occur in bound-dissociative transitions with \ce{He+} as electron acceptor: the model predicts more pronounced oscillations and shows a faster decay with increasing KER compared to the R-matrix results. 
These differences likely arise from limitations of the model, including the neglect of higher-order multipole terms and interference between energy and electron transfer mechanisms. 
The reliability of the R-matrix approach, on the other hand, is affected by the use of the AN approximation in B to X and B to A, which becomes less accurate at higher KER.
%interference between energy and electron transfer mechanisms, and use of the AN approximation in B to X and B to A, which becomes less accurate at higher KER.

%\input{conclusion}
\section{Conclusion}
\label{sec:conclusion}

We have presented a comprehensive theoretical investigation of how nuclear motion influences the ICEC cross section in the positively charged helium-neon dimer -- an example of a weakly bound system.
In our approach, we incorporate vibrational dynamics into an analytical model of ICEC, including both the energy and electron transfer mechanisms in terms of a semi-empirical model.
To validate the model, we implemented an approximation based on fixed-nuclei \textit{ab-initio} R-matrix results, finding agreement within an order of magnitude.  
The model not only reproduces the R-matrix cross sections but also provides valuable interpretational insight.

Our results demonstrate that including nuclear motion and temperature dependence leads to important modifications of the ICEC cross sections. 
These include changes in threshold position, threshold behavior, energy dependence, outgoing electron energy, and overall magnitude.
When the neon cation captures the incoming electron, the cross section rises near the threshold and is predominantly followed by dissociation at higher electron energies -- a tendency that becomes even more pronounced at elevated temperatures.
In contrast, electron capture by the helium cation favors bound-bound transitions, with slightly reduced cross sections observed at lower temperatures. 
We find that electron transfer dominates the cross section, as the bound vibrational states are localized in the short-range region where the asymptotic energy-transfer approximation breaks down.
Although nuclear motion can slightly reduce the total cross section compared to fixed-nuclei calculations at equilibrium geometries, ICEC remains several orders of magnitude more efficient in the helium-neon dimer than photorecombination, reinforcing its physical relevance.

It is important to note that, although we treat the electronic states X and A as distinct within our theoretical framework, such separation may not be feasible experimentally due to limited resolution, which could obscure individual contributions.
Additionally, while our treatment neglects molecular rotations, their inclusion would introduce further fine structure in the bound-bound transitions. However, resolving this substructure would be experimentally challenging, as the energy splitting between rotational states within the same vibrational level is minimal.

The influence of nuclear dynamics on ICEC has important implications for both theoretical modeling and experimental detection. Our results highlight the need for including this dynamics into ICEC descriptions, particularly in molecular and weakly bound systems.

\section*{Supplementary information}
We present in the supplementary material the ICEC cross sections for higher initial vibrational states, which underlie the temperature dependence discussed in Sec.~\ref{sec:temperature_dependence}.
We further show the temperature-dependent cross sections for all electronic transitions, as well as the spectra for higher initial vibrational states.

\section*{Acknowledgements}
We thank the DFG-ANR for financial support through the QD4ICEC project.
The corresponding grant numbers are FA 1989/1-1 and ANR-22-CE92-0071-01.
E. F. and E. M. J. thank for a travel grant provided by the
Zentrum für Frankophone Welten at the University of Tübingen,
E. F. furthermore acknowledges funding by LISA$^+$ at the University of Tübingen. 
J. P. D. acknowledges financial support from the COST Action CA18212–Molecular Dynamics in the GAS phase (MD-GAS), supported by COST (European Cooperation in Science and Technology).

\section*{Author statements}
\subsection*{Conflict of interest}
The authors have no conflicts to disclose.

\subsection*{Author Contributions}

\textbf{Conceptualization}: E.M.J., J.\v{S}., N.S., E.F.
\textbf{Formal Analysis}: E.M.J., J.\v{S}.
\textbf{Funding}: N.S., E.F.
\textbf{Investigation}: E.M.J., J.\v{S}.
\textbf{Methodology}: E.M.J., J.\v{S}., J.P.D., P.K., N.S., E.F.
\textbf{Software}: E.M.J., J.\v{S}.
\textbf{Supervision}: P.K., N.S., E.F.
\textbf{Visualization}: E.M.J., J.\v{S}.
\textbf{Writing -- original draft}: E.M.J. (equal), J.\v{S}. (equal).
\textbf{Writing -- review \& editing}: E.M.J., J.\v{S}., J.P.D., P.K., N.S., E.F.

\section*{Data availability}
The data that support the findings of this study are openly available and cited at the appropriate locations within this paper.

%\bibliography{refs}

\end{document}